\documentclass{ar-1col}
\usepackage[numbers]{natbib}
\usepackage{url, bm, amsmath, amssymb}
\newcommand{\bP}{{\bf p}}
\newcommand{\bJ}{{\bf J}}
\newcommand{\bF}{{\bf F}}
\newcommand{\tobs}{t}

\jname{Annu. Rev. Condens. Matter Phys.}
\title{Irreversibility and biased ensembles in active matter: Insights from stochastic thermodynamics}

\author{\'Etienne Fodor,$^1$  Robert L. Jack,$^{2,3}$\\and Michael E. Cates$^2$
	\affil{$^1$Department of Physics and Materials Science, University of Luxembourg, L-1511 Luxembourg; email: etienne.fodor@uni.lu}
	\affil{$^2$Department of Applied Mathematics and Theoretical Physics, University of Cambridge, Wilberforce Road, Cambridge CB3 0WA, United Kingdom; email: rlj22@cam.ac.uk, m.e.cates@damtp.cam.ac.uk}
	\affil{$^3$Yusuf Hamied Department of Chemistry, University of Cambridge, Lensfield Road, Cambridge CB2 1EW, United Kingdom}
}
\begin{document}

\begin{abstract}
	Active systems evade the rules of equilibrium thermodynamics by constantly dissipating energy at the level of their microscopic components. This energy flux stems from the conversion of a fuel, present in the environment, into sustained individual motion. It can lead to collective effects without any equilibrium equivalent, such as a phase separation for purely repulsive particles, or a collective motion (flocking) for aligning particles. Some of these effects can be rationalized by using equilibrium tools to recapitulate nonequilibrium transitions. An important challenge is then to delineate systematically to which extent the character of these active transitions is genuinely distinct from equilibrium analogs. We review recent works that use stochastic thermodynamics tools to identify, for active systems, a measure of irreversibility comprising a coarse-grained or {\it informatic} entropy production. We describe how this relates to the underlying energy dissipation or {\it thermodynamic} entropy production, and how it is influenced by collective behavior. Then, we review the possibility to construct thermodynamic ensembles out-of-equilibrium, where trajectories are biased towards atypical values of nonequilibrium observables. We show that this is a generic route to discovering unexpected phase transitions in active matter systems, which can also inform their design.
\end{abstract}


\maketitle

% ===============================================================================

\section{INTRODUCTION}

Active matter is a class of nonequilibrium systems whose components extract energy from the environment to produce an autonomous motion~\cite{Marchetti2013, Bechinger2016, Marchetti2018}. Examples are found in biological systems, such as swarms of bacteria~\cite{Elgeti2015} and assemblies of cells~\cite{Ladoux2017}; social systems, such as groups of animals~\cite{Cavagna2014} and human crowds~\cite{Bartolo2019}; synthetic systems, such as vibrated polar particles~\cite{Dauchot2010} and self-catalytic colloids~\cite{Palacci2013}. The combination of individual self-propulsion and interactions between individuals can lead to collective effects without any equivalent in equilibrium. Examples include collective directed motion, as observed in bird flocks~\cite{Cavagna2014}, and the spontaneous formation of clusters made of purely repulsive particles, as reported for Janus colloids in a fuel bath~\cite{Palacci2013}. To study these effects, minimal models have been proposed based on simple dynamical rules. Some are formulated at the level of individual particles, for instance with automaton rules~\cite{Vicsek1995} or by extending  Langevin dynamics~\cite{Fily2012}. Others describe the system at coarse-grained level in terms of hydrodynamic fields~\cite{Toner1995, Wittkowski2014}. The latter can be obtained systematically by coarse-graining the particle dynamics or postulated phenomenologically. Both the microscopic and hydrodynamic approaches have successfully reproduced experimental behavior, such as the emergence of a long-ranged polar order, known as the flocking transition~\cite{Chate2020}, and phase separation that occurs without any microscopic attraction, known as motility-induced phase separation (MIPS)~\cite{Cates2015}.

Although active systems evade the rules of equilibrium statistical mechanics, some works have built a framework to predict their properties based on the partial applicability of thermodynamic concepts beyond equilibrium. A first approach was to map some active systems onto equilibrium ones with a similar steady state, allowing one to define effective free-energies for active matter~\cite{Tailleur2008, Maggi2015}. Other studies have extended the definitions of standard observables including pressure~\cite{Marchetti2014, Brady2014, Solon2015}, surface tension~\cite{Speck2015, Zakine2020}, and chemical potential~\cite{Paliwal2018, Guioth2019}, hoping to establish equations of state relating, for instance, pressure and density. Interestingly, one finds that the existence of such state functions cannot generically be relied upon ({\it e.g.}, the pressure on a wall can depend on the type of wall)~\cite{Solon2015}, highlighting the limitations of equilibrium analogies. In trying to build a thermodynamic framework for active matter, an important challenge is then to identify regimes where it is possible to deploy equilibrium tools, and to clearly distinguish these from {\it genuine} nonequilibrium regimes. In other words, how should we delineate where and when activity really matters in the emerging phenomenology? And, most importantly, can we define a systematic, unambiguous measure of the departure from equilibrium?

In passive systems, the steady state (Boltzmann) distribution involves the Hamiltonian which also drives the micro-dynamics. This ensures \emph{thermodynamic consistency} of the dynamics, which further implies that the mechanical and thermodynamic definitions of pressure are equivalent, and precludes (real space) steady-state currents. While such currents offer a clear nonequilibrium signature of activity, as observed for instance in collective motion~\cite{Chate2020}, it is more challenging to distinguish, say, MIPS from standard phase separation without tracking individual particle motion~\cite{Cates2015}. In particular it is not helpful to define departure from equilibrium {\it via} deviation from an effective ``Boltzmann distribution'' in systems that are not thermodynamically consistent. Instead, the cornerstone of modern statistical mechanics, which allows one to dissociate fundamentally active and passive systems, is the reversibility of equilibrium dynamics~\cite{Onsager1931}. In a steady equilibrium state, forward and backward dynamics are indistinguishable, since all fluctuations exhibit time-reversal symmetry (TRS). This constraint entails other important properties, such as the fluctuation-dissipation theorem (FDT)~\cite{Kubo1966}, and the absence of dissipated heat at equilibrium. For active systems, the irreversibility of the dynamics then stands out as the key differentiating property that causes the violation of these and other equilibrium laws.

To quantify the breakdown of TRS, we use stochastic thermodynamics~\cite{Sekimoto1998, Seifert2012}. This framework extends standard notions of thermodynamics, such as the first and second laws, to fluctuating trajectories. It was first developed for systems in contact with one or more reservoirs, such as energy and/or volume reservoirs, provided that each reservoir satisfies some equilibrium constraints. Such constraints do not preclude the system to operate out-of-equilibrium, for instance under external fields and/or thermostats at different temperatures: In this context, the thermodynamic consistency of the dynamics yields some explicit connections between irreversibility, dissipated heat and entropy production. In contrast, the dynamics of active systems are often formulated {\it via} phenomenological arguments, which do not follow {\it a priori} the requirements of thermodynamic consistency. A natural question is then: Is it legitimate to extend the methods of stochastic thermodynamics to active matter? And, what can we learn from its irreversibility measures when their connection with other thermodynamic observables becomes blurred?

Another interesting direction in active matter theory is to generalize familiar thermodynamic ensembles to nonequilibrium settings.  Here the important role of steady-state currents and other types of irreversible dynamics make it insufficient to study ensembles of configurations (characterized by a stationary measure, of which the Boltzmann distribution is an example). Instead we must address ensembles of trajectories, as used previously to analyze fluctuation theorems~\cite{Maes1999gibbs}, and other dynamical effects~\cite{Derrida2007,Lecomte2007,Garrahan2007}. Such ensembles are built similarly to the canonical ensemble of statistical mechanics, with relevant (extensive) observables coupled to conjugate (intensive) fields. In practice, one selects a dynamical observable of interest and focusses on dynamical trajectories where it has some atypical target value.  The resulting {\it biased ensembles} of rare trajectories, are intrinsically connected to {\em large deviation theory}~\cite{Touchette2009,Jack2020}. As such, they provide insight into mechanisms for unusual and interesting fluctuations. In the passive context, they have proven useful for unveiling dynamical transitions, {\em e.g.}, in kinetically constrained models of glasses~\cite{Garrahan2007}. This leads one to ask: How do the transitions in biased ensembles of active matter differ from their passive counterparts? And then, can we identify settings in which bias-induced transitions emerge in active matter that have no passive equivalent?

In what follows (Section~\ref{sec:epr}), we describe how the irreversibility of active systems can be quantified, both for particle-based and hydrodynamic theories, using generalized forms of entropy production. We discuss how to relate these, in certain cases, to energy dissipation, and how they allow to identify phases and/or spatial regions where activity comes to the fore. In Section~\ref{sec:bias}, we then describe how biased ensembles provide novel insights on the emergence of collective effects in active matter. We review some useful tools of large deviation theory, such as representing dynamical bias in terms of control forces, show the utility of these methods in an active context, and discuss how they can reveal unexpected transitions in generic active systems. In Section~\ref{sec:conc} we give a brief conclusion that includes implications of these methods for the rational design of new materials.

% ===============================================================================

\section{IRREVERSIBILITY AND DISSIPATION}\label{sec:epr}

\subsection{Modeling active matter: From particles to hydrodynamic fields}\label{sec:model}

A typical particle-based dynamics starts from the seminal Langevin equation. Provided that inertia is negligible, this balances the forces stemming from the heat bath (damping and thermal noise), the force deriving from a potential $U$, and non-conservative forces ${\bf f}_i$:
\begin{equation}\label{eq:dyn}
	\dot{\bf r}_i = \mu \big( {\bf f}_i - \nabla_i U) + \sqrt{2\mu T} {\bm\xi}_i ,
\end{equation}
where $\mu$ is the mobility, $T$ the temperature of the bath, and ${\bm\xi}_i$ a set of zero-mean, unit-variance white Gaussian noises: $\langle\xi_{i\alpha}(t)\xi_{j\beta}(t')\rangle = \delta_{ij}\delta_{\alpha\beta}\delta(t-t')$, where Latin and Greek indices respectively refer to particle labels and spatial coordinates. Hereafter, we refer to such noise as {\it unit white noise}. The potential $U$ describes particle interactions and/or an external perturbation applied by the operator. The non-conservative forces ${\bf f}_i$ model self-propulsion, which converts a source of energy, present in the environment, into directed motion. In principle, ${\bf f}_i$ can include external perturbations that do not derive from a potential. When ${\bf f}_i={\bf 0}$, the system is at equilibrium with Boltzmann statistics ($\propto e^{-U/T}$).

The self-propulsion force ${\bf f}_i$ is often modeled as effectively an additional noise. In contrast with the thermal noise ${\bm\xi}_i$, it has some {\it persistence} which captures the propensity of active particles to sustain directed motion. Typically, a persistence time $\tau$ sets the exponential decay of the two-point correlations: $\langle f_{i\alpha}(t)f_{j\beta}(0) \rangle = \delta_{ij}\delta_{\alpha\beta} f_0^2 {\rm e}^{-|t|/\tau}$. In recent years, two variants have emerged as popular models of self-propelled particles. First, for Active Ornstein-Uhlenbeck Particles (AOUPs), the statistics of ${\bf f}_i$ is Gaussian, so that it can be viewed as obeying an (autonomous) Ornstein-Uhlenbeck process~\cite{Maggi2015, Nardini2016}:
\begin{equation}\label{eq:aoup}
	{\rm AOUP} : \quad \tau\dot{\bf f}_i = - {\bf f}_i + f_0 \sqrt{2\tau} {\bm\zeta}_i ,
\end{equation}
where the ${\bm\zeta}_i$ are unit white noises. Scaling the amplitude as $\mu f_0 = \sqrt{\mu T_a/\tau}$, the self-propulsion itself converges to a white noise source in the limit of vanishing persistence, in which regime the system reduces to passive particles at temperature $T+T_a$. Second, for Active Brownian Particles (ABPs), the magnitude of ${\bf f}_i$ is now fixed, which leads to non-Gaussian statistics, described by the following process in two dimensions~\cite{Fily2012}:
\begin{equation}\label{eq:abp}
	{\rm ABP} : \quad {\bf f}_i = f_0 (\cos\theta_i, \sin\theta_i) ,
	\quad
	\dot\theta_i = \sqrt{2/\tau} \, \eta_i ,
\end{equation}
where $\eta_i$ is a unit white noise. In Equations~\ref{eq:aoup} and~\ref{eq:abp}, the self-propulsion is independent for each particle. This corresponds to isotropic particles, which undergo MIPS for sufficiently large persistence and density~\cite{Cates2015}. In contrast, many models consider alignment interactions, yielding a collective (oriented) motion at high density and enhanced persistence~\cite{Chate2020}.

To study the emergence of nonequilibrium collective effects, it is helpful to introduce the relevant hydrodynamic fields. These can be identified, and their dynamical equations found, by coarse-graining the microscopic equations of motion, using standard tools of stochastic calculus~\cite{Dean1996}. Another approach consists in postulating such hydrodynamic theories from phenomenological arguments, respecting any conservation laws and spontaneously broken symmetries~\cite{Marchetti2013}. For instance, a theory of MIPS can be found by considering the dynamics of a scalar field $\phi({\bf r},t)$ that encodes the local concentration of particles~\cite{Wittkowski2014, Tjhung2018}:
\begin{equation}\label{eq:dyn_phi}
	\dot\phi = -\nabla\cdot{\bf J} = - \nabla\cdot \bigg(- \lambda \nabla \frac{\delta\cal F}{\delta\phi} + {\bf J}_\phi + \sqrt{2\lambda D}{\bm\Lambda}_\phi \bigg) ,
\end{equation}
where $\lambda$ is collective mobility, $D$ a noise temperature, and ${\bm\Lambda}_\phi$  (spatiotemporal) unit white noise. The term ${\bf J}_\phi$ embodies contributions driving the current ${\bf J}$ of $\phi$ that cannot be derived from any ``free energy'' functional ${\cal F}$ as $-\lambda\nabla(\delta {\cal F}/\delta\phi)$. For instance, choosing a standard $\phi^4$, square-gradient functional for $\cal F$ captures phase separation in equilibrium. On introducing activity, symmetry arguments enforce that ${\bf J}_\phi$ takes the form $\nabla(\nabla\phi)^2$ and/or $(\nabla\phi)(\nabla^2\phi)$ to lowest order in $\phi$ and its gradients~\cite{Wittkowski2014, Tjhung2018}. The former term shifts the coexisting densities with respect to the equilibrium model~\cite{Wittkowski2014}, whereas the latter can lead to microphase separation, with either the vapour phase decorated with liquid droplets, or the liquid phase decorated with vapour bubbles~\cite{Tjhung2018}. Note that, although the conservative term $-\lambda\nabla(\delta {\cal F}/\delta\phi)$ captures all the contributions to the coarse-grained dynamics that respect TRS, it may be depend on nonequilibrium parameters of the microscopic dynamics~\cite{Tjhung2018}.

The dynamics in Equation~\ref{eq:dyn_phi} can also be coupled to a polar field ${\bf p}({\bf r},t)$, representing the local mean orientation of particles, to study for instance collective {\it flocking} motion~\cite{Chate2020}:
\begin{equation}\label{eq:dyn_p}
	\dot{\bf p} =  - \frac{\delta\cal F}{\delta\bf p} + {\bf J}_p + \sqrt{2 D} {\bm\Lambda}_p ,
\end{equation}
where the rotational mobility is taken as unity, and ${\bm\Lambda}_p$ is another spatiotemporal unit white noise. As with ${\bf J}_\phi$, the  term ${\bf J}_p$ represents active relaxations that cannot be written as free-energy derivatives. Note that, for the passive limit of the coupled model to respect detailed balance, the free energy $\cal F$ in Equation~\ref{eq:dyn_p}  must be the same as that in Equation~\ref{eq:dyn_phi}, and the value of $D$ in each equation must also coincide. In the absence of activity (${\bf J}_\phi={\bf J}_p={\bf 0}$), the system has Boltzmann statistics $\propto e^{-{\cal F}/D}$. To capture the emergence of polar order, a minimal choice is to take $\cal F$ as a ${\bf p}^4$ functional, with ${\bf J}_\phi$ proportional to ${\bf p}$, and ${\bf J}_p$ a linear combination of $\nabla\phi$, $\phi{\bf p}$ and $({\bf p}\cdot\nabla){\bf p}$~\cite{Marchetti2013, Chate2020}. This can yield polar bands traveling across an apolar background, which is a key signature of aligning active particles models~\cite{Vicsek1995}. In fact, much of the physics of flocks is retained by setting $\lambda=0$ in Equation~\ref{eq:dyn_phi}~\cite{Toner1995}. On the other hand, more complicated theories, describing the emergence of nematic order~\cite{Chate2020} and/or the coupling to a momentum conserving fluid~\cite{Tiribocchi2015, Nardini2017}, introduce additional hydrodynamic fields beyond $\phi$ and ${\bf p}$ as considered above.

% -------------------------------------------------------------------------------

\subsection{How far from equilibrium is active matter?}\label{sec:path}

\subsubsection{Forward and time-reversed dynamics: Path-probability representation}

\begin{figure}
	\includegraphics[width=7cm]{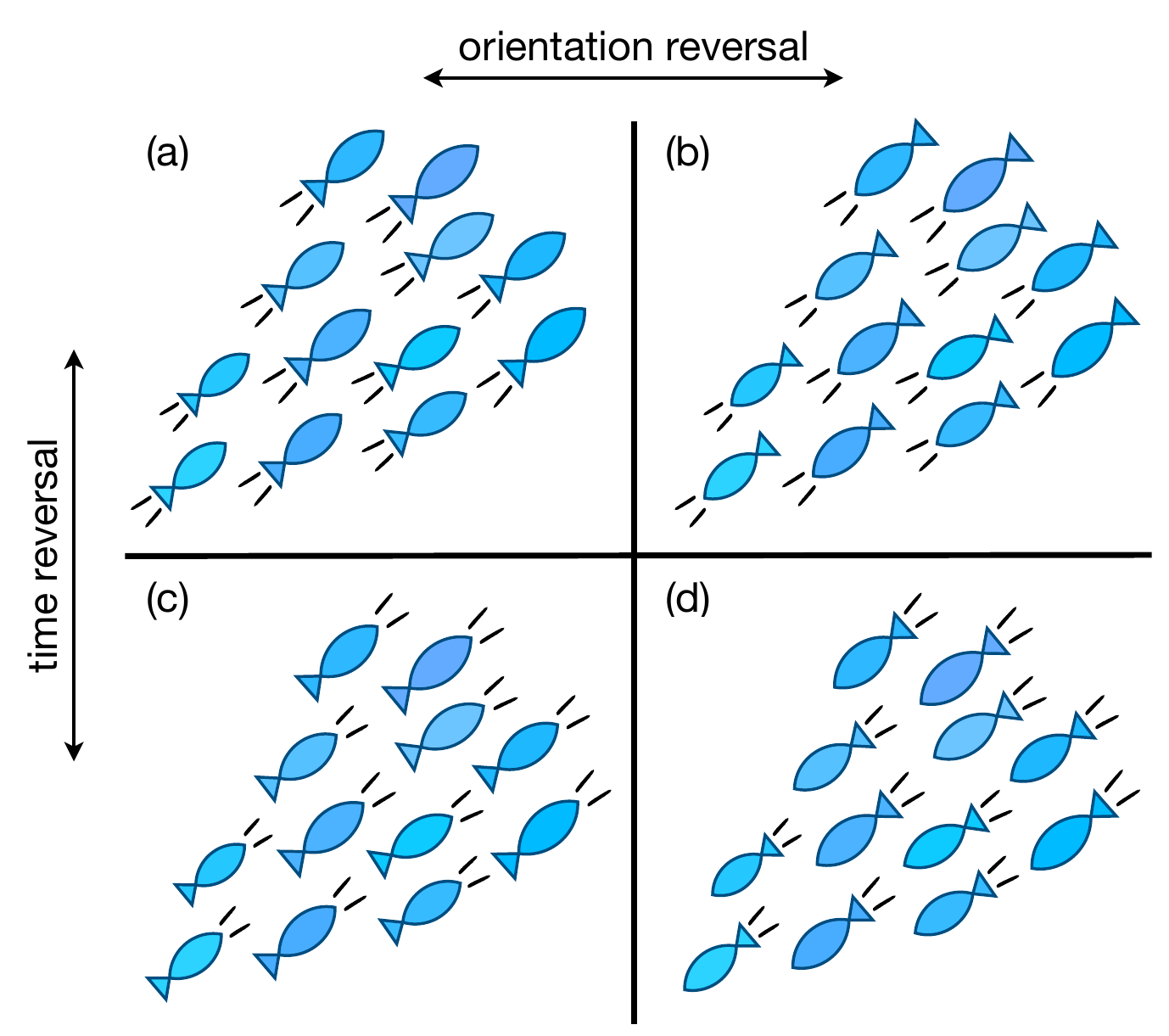}
	\caption{
Time reversal (TR) for active particles (fish). Each has an orientation (head-tail axis) and moves (swims) as indicated by the {\it wake lines} behind it. The TR operation may or may not reverse orientations. (a) Natural dynamics. (b) Reversing orientation but not motion: fish swim tail-first along their original directions. (c) Reversing motion but not orientation: fish swim tail-first, opposite to their original direction. (d) Reversing both motion and orientation: fish swim head-first, opposite to their original direction. The situations in (b,c) occur with extremely low probability in the natural dynamics (requiring exceptional noise realizations). In dilute regimes (not shown) case (d) is equiprobable to (a): individual fish swim head-first in both cases. For the shoals shown here, however, (d) is less probable than (a), because the natural dynamics has more fish at the front of the shoal: TRS is broken at a collective level even if orientations are reversed.
}
	\label{fig1}
\end{figure}

To quantify the irreversibility of active dynamics, we compare the path probability $\cal P$ to realize a dynamical trajectory (across a time interval $[0,t]$) with that for its time-reversed counterpart~\cite{Sekimoto1998, Seifert2012, Lebowitz1999}. As we shall see, choosing the time-reversed dynamics is a matter of subtlety~\cite{Nardini2016, Nardini2017, Mandal2017, Puglisi2017, Speck2018, Caprini2018, Shankar2018, Seifert2018, Ramaswamy2018, Bo2019, Borthne2020}. For the particle-based dynamics in Equation~\ref{eq:dyn}, exploiting the known statistics of the (unit white) noise, trajectories of position ${\bf r}_i$ and self-propulsion ${\bf f}_i$ have probability ${\cal P} \sim e^{-\cal A}$~\cite{Onsager1953}, with action
\begin{equation}\label{eq:action}
	{\cal A} = \frac{1}{4\mu T} \int_0^t \sum_i \Big[ \dot{\bf r}_i + \mu \big(\nabla_i U - {\bf f}_i\big) \Big]^2 dt' + {\cal A}_f .
\end{equation}
We use Stratonovitch discretization, omitting a time-symmetric contribution that is irrelevant in what follows. ${\cal A}_f$ specifies the statistics of the self-propulsion force ${\bf f}_i$.  For AOUPs and ABPs, ${\bf f}_i$ has autonomous dynamics, independent of position ${\bf r}_i$.  One can then in principle integrate out that dynamics to obtain a reduced action ${\cal A}_r$. So far, the explicit expression has been derived only for AOUPs~\cite{Nardini2016, Bo2019, Caprini2019, Martin2020b}:
\begin{equation}\label{eq:action_red}
	{\cal A}_r = \int_0^t dt' \int_0^t dt'' \sum_i \big( \dot{\bf r}_i + \mu\nabla_i U \big) \big|_{t'} \cdot \big( \dot{\bf r}_i + \mu\nabla_i U \big)\big|_{t''} \Gamma(t'-t'') .
\end{equation}
The kernel $\Gamma$ is an even function of time $t$, which depends on $\mu$, $T$, $f_0$, and $\tau$. It reduces to $\Gamma(t) = \delta(t) / (4\mu(T+T_a))$ for vanishing persistence ($\tau=0$), when the self-propulsion amplitude scales as $\mu f_0=\sqrt{\mu T_a/\tau}$~\cite{Bo2019}, as expected for passive particles. The actions $\cal A$ and ${\cal A}_r$ describe the same dynamics from different viewpoints. The former tracks the realizations of position {\it and} self-propulsion, whereas the latter follows the time evolution of positions only. Thus ${\cal A}_r$ treats the self-propulsion as a source of noise, without resolving its time evolution. In contrast, $\cal A$ regards ${\bf f}_i$ as a configurational coordinate, which might in principle be read out from the particle shape. For instance, Janus particles, which are spheres with distinct surface chemistry on two hemispheres~\cite{Bechinger2016, Palacci2013}, generally self-propel along a body-fixed {\it heading vector} pointing from one to the other.  See also {\bf Figure~\ref{fig1}}.

To compare forward and backward trajectories, one needs to define a time-reversed version of the dynamics. This necessarily transforms the time variable as $t'\to t-t'$, and it can also involve flipping some of the variables/fields (as detailed below). For particle-based dynamics, time reversal of the positions ${\bf r}_i$ is unambiguous.  Writing $\cal P$  in terms of the reduced action ${\cal A}_r$ in Equation~\ref{eq:action_red}, the time-reversed counterpart ${\cal A}^R_r$ reads
\begin{equation}\label{eq:action_red_rev}
	{\cal A}^R_r = \int_0^t dt' \int_0^t dt'' \sum_i \big( \dot{\bf r}_i - \mu\nabla_i U \big) \big|_{t'} \cdot \big( \dot{\bf r}_i - \mu\nabla_i U \big)\big|_{t''} \Gamma(t'-t'') \,.
\end{equation}
Here we have changed variables as $\{t',t''\}\to\{t-t',t-t''\}$, and used that $\Gamma$ is even. For the full action $\cal A$, defining the time-reversed dynamics requires that we fix the time signature of the self-propulsion ${\bf f}_i$. Indeed, there are two possible reversed actions, ${\cal A}^{R,\pm}$, where $+$ and $-$ indices refer respectively to even (${\bf f}_i\to{\bf f}_i$) and odd (${\bf f}_i\to-{\bf f}_i$) self-propulsion:
\begin{equation}\label{eq:action_rev}
	{\cal A}^{R,\pm} = \frac{1}{4\mu T} \int_0^t \sum_i \Big[ \dot{\bf r}_i - \mu\big(\nabla_i U \mp {\bf f}_i \big) \Big]^2 dt' + {\cal A}_f .
\end{equation}
We assumed that ${\cal A}_f$ is the same for forward and time-reversed dynamics which holds for ABPs/AOUPs (but not for systems with aligning interactions~\cite{Spinney2018}). ${\cal A}^R_r$ and ${\cal A}^{R,\pm}$ correspond to three different definitional choices of the time-reversed dynamics. The first of these deliberately discards the realizations of the self-propulsion forces, in order to compare the forward and backward motion of particle positions. This choice is widely used for AOUPs, where particles have no heading vector. It is also the only possible choice when $T=0$ in Equation~\ref{eq:dyn}, see~\cite{Nardini2016}. In contrast, ${\cal A}^{R,\pm}$, which each assume that both position and self-propulsion are tracked, are usually chosen for particles with a heading vector~\cite{Speck2016, Shankar2018, Spinney2018}.

For the field theories of Section~\ref{sec:model}, defining the time-reversed dynamics requires us to choose the time signature of each field. This choice depends on the physical system and the phases under scrutiny (Section~\ref{sec:coll}). Scalar fields associated with a local density are clearly even~\cite{Wittkowski2014, Nardini2017}, whereas scalar fields schematically representing polarization~\cite{Solon2013}, or stream functions for fluid flow~\cite{Alert2020}, should generally be odd. Similarly, the vector field $\bf p$ can be chosen even~\cite{Markovich2020} or odd~\cite{Ramaswamy2018}, depending on whether  it is viewed as the local orientation of particles (mean heading vector, see {\bf Figure~\ref{fig1}}) or directly as a local velocity~\cite{Toner1995}. One also has to decide which fields to retain or ignore when comparing forward and backward paths. This choice is the counterpart at field level of retaining (${\cal A}$) or ignoring (${\cal A}_r$) the propulsive forces in a system of AOUPs. For example, in a system with $\phi$ and ${\bf p}$ variables, one can choose whether or not to keep separate track of the density current ${\bf J}$ alongside $\phi$ and ${\bf p}$.

In general, there are four different versions for time-reversed dynamics of Equations~\ref{eq:dyn_phi} and~\ref{eq:dyn_p}, provided that each one of the fields $\phi$ and $\bf p$ can be either odd or even. Often, though, constraints on the time signatures of the fields restrict these choices. To ensure that the dynamics is invariant under time reversal in the passive limit, the free energy $\cal F$ must be the same in forward and time-reversed dynamics. This enforces that odd fields can only appear as even powers in $\cal F$. Hence if ${\bf p}$ appears linearly in $\cal F$, which for liquid crystal models often includes a ${\bf p}\cdot\nabla\phi$ or {\it anchoring} term, it must be chosen even. Further constraints arise if some of the possible noise terms are set to zero ({\it e.g.},~\cite{Toner1995}) so that certain fields are deterministically enslaved to others. For instance if $\dot\phi = - \nabla\cdot{\bf p}$ (without noise), for consistency $\phi$ and $\bf p$ necessarily have different signatures.

% -------------------------------------------------------------------------------

\subsubsection{Distance from equilibrium: Breakdown of time-reversal symmetry}

For a given dynamics, once its time-reversed counterpart has been chosen, we can define systematically an irreversibility measure $\cal S$ as~\cite{Sekimoto1998, Seifert2012, Lebowitz1999}
\begin{equation}\label{eq:epr}
	{\cal S} = \underset{t\to\infty}{\lim} \frac{1}{t} \bigg\langle \ln \frac{\cal P}{{\cal P}^R} \bigg\rangle = \underset{t\to\infty}{\lim} \frac{1}{t} \big\langle{\cal A}^R - \cal A\big\rangle .
\end{equation}
The average $\langle\cdot\rangle$ is taken over noise realizations. The limit of long trajectories gets rid of any transient relaxations to focus on steady-state fluctuations. Equation~\ref{eq:epr} is a cornerstone of stochastic thermodynamics~\cite{Sekimoto1998, Seifert2012, Lebowitz1999}. It was first established in thermodynamically consistent models~\cite{Lebowitz1999}, where $\cal S$ can be shown to be the entropy production rate (EPR) governing the dissipated heat~\cite{Sekimoto1998, Seifert2012}. In models where $\cal S$ has this meaning, thermodynamics constrains the choice of time reversal. In Section~\ref{sec:ener}, we return to the question of how far the thermodynamic interpretation extends to active systems. Meanwhile, $\cal S$ already offers an unambiguous measure of TRS breakdown in active matter, which we refer to as {\it informatic} EPR (IEPR). Since $\cal S$ changes when the dynamics is coarse-grained, by eliminating unwanted variables, it differs when retaining ($\cal A$) or ignoring (${\cal A}_r$) the propulsive forces.

In particle-based dynamics, choosing the action ${\cal A}_r$ so that only particle positions are tracked, the corresponding IEPR ${\cal S}_r$ follows from Equations~\ref{eq:action_red},~\ref{eq:action_red_rev} and~\ref{eq:epr} as
\begin{equation}\label{eq:epr_r}
	{\cal S}_r = - 4 \mu \int_{-\infty}^\infty \Gamma(t) \sum_i \big\langle \dot{\bf r}_i(t)\cdot\nabla_i U(0) \big\rangle dt ,
\end{equation}
where we have again used that $\Gamma$ is even. The IEPR ${\cal S}_r$ vanishes in the absence of any potential, showing that the dynamics satisfies TRS for free active particles, and also for an external harmonic potential $U\sim{\bf r}_i^2$~\cite{Nardini2016, Martin2020b}. When $T=0$, it reduces to ${\cal S}_r = \tau \langle (\sum_i \dot{\bf r}_i\cdot\nabla_i)^3 U \rangle /(2(\mu f_0)^2)$~\cite{Nardini2016}. When (by using the full action $\cal A$) one follows the dynamics of both position and self-propulsion, the two possible IEPRs found from Equations~\ref{eq:action},~\ref{eq:action_rev} and~\ref{eq:epr}, are
\begin{equation}
	\label{eq:epr+}
	{\cal S}^+ = \frac{1}{T} \sum_i \big\langle \dot{\bf r}_i\cdot{\bf f}_i \big\rangle ,
	\qquad
	{\cal S}^- = \frac{\mu}{T} \sum_i \big\langle \nabla_i U\cdot{\bf f}_i \big\rangle .
\end{equation}
We have used that $\frac{1}{t}\int_0^t \sum_i \dot{\bf r}_i\cdot\nabla_i U dt' = \frac{U(t)-U(0)}{t}$ vanishes at large $t$. Substituting in Equation~\ref{eq:epr+} the expression for $\dot{\bf r}_i$ from Equation~\ref{eq:dyn} yields ${\cal S}^+ + {\cal S}^- = N\mu f_0^2/T$, where $N$ is the particle number, using that self-propulsion ${\bf f}_i$ and thermal noise $\sqrt{2\mu T}{\bm\xi}_i$ are uncorrelated. Therefore, in the absence of any potential $U$, ${\cal S}^-$ vanishes identically whereas ${\cal S}^+$ remains non-zero. Indeed, ${\cal S}^-$ compares trajectories whose velocity and self-propulsion both flip on time reversal, retaining alignment (up to thermal noise) between the two: The forward and backward dynamics are indistinguishable unless potential forces intervene. In contrast, ${\cal S}^+$ quantifies how different trajectories are when particles move either along with, or opposite to, their self-propulsion ({\bf Figure~\ref{fig1}}), yielding the contribution $N\mu f_0^2/T$ even when $U=0$.

The IEPR ${\cal S}^+$ is lowest (and ${\cal S}^-$ highest) when the propulsive force ${\bf f}_i$ balances the interaction force $-\nabla_i U$ so that particles are almost arrested. Hence, both IEPRs are sensitive to the formation of particle clusters, which is associated with dynamical slowing-down for isotropic particles~\cite{Cates2015}, and to formation of a polarized state for aligning particles~\cite{Chate2020}. Note that ${\cal S}^+$ is proportional to the contribution of self-propulsion to pressure, known as swim pressure~\cite{Brady2014, Solon2015}. Finally, in the presence of alignment, any dynamical interaction torques appear explicitly in  $\cal S$ ~\cite{Spinney2018}. We defer further discussion on how interactions shape irreversibility, for both particle-based dynamics and field theories, to Section~\ref{sec:coll}.

% -------------------------------------------------------------------------------

\subsection{Energetics far from equilibrium: Extracted work and dissipated heat}\label{sec:ener}

\subsubsection{Particle-based approach: Energy transfers in microscopic dynamics}

A major success of stochastic thermodynamics is to extend the definition of observables from classical thermodynamics to cases where fluctuations cannot be neglected~\cite{Sekimoto1998, Seifert2012, Lebowitz1999}. Although first proposed for thermodynamically consistent models, this approach can be extended to some (not all) types of active matter. Considering that the potential $U$ depends on the set of control parameters $\alpha_n$, the work $\cal W$ produced by varying $\alpha_n$ during a time $t$ reads~\cite{Sekimoto1998, Seifert2012}
\begin{equation}\label{eq:work}
	{\cal W} = \int_0^t \sum_n \dot\alpha_n \frac{\partial U}{\partial\alpha_n} dt' .
\end{equation}
Note that $\cal W$ is stochastic due to the thermal noise and the self-propulsion, even though the protocol $\alpha_n(t)$ is deterministic. Equation~\ref{eq:work} relies on the precept that some external operator perturbs the system though $U$, without prior knowledge of the detailed dynamics; it applies equally for active and passive systems.

The heat $\cal Q$ is now defined as the energy delivered by the system to the surrounding thermostat. For the dynamics of Equation~\ref{eq:dyn}, the effect of the thermostat is encoded in the damping force and the thermal noise; the fluctuating observable $\cal Q$ then follows as~\cite{Sekimoto1998, Seifert2012}
\begin{equation}\label{eq:heat}
	{\cal Q} = \int_0^t \sum_i \frac{\dot{\bf r}_i}{\mu} \cdot \big( \dot{\bf r}_i - \sqrt{2\mu T}{\bm\xi}_i \big) dt' .
\end{equation}
Substituting Equation~\ref{eq:dyn} into Equation~\ref{eq:heat}, and using the chain rule $\dot U = \sum_n \dot\alpha_n (\partial U/\partial\alpha_n)+ \sum_i \dot{\bf r}_i\cdot\nabla_iU$, gives a relation between energy $U$, work $\cal W$, and heat $\cal Q$:
\begin{equation}\label{eq:flt}
	U(t) - U(0) = {\cal W} - {\cal Q} + \int_0^t \sum_i \dot{\bf r}_i\cdot{\bf f}_i dt' .
\end{equation}
In passive systems, the non-conservative force ${\bf f}_i$ represents some intervention by the external operator (beyond changes in $U$), so that the term $\bar{\cal  W} \equiv \int_0^t \sum_i \dot{\bf r}_i\cdot{\bf f}_i dt'$ can be absorbed into the work $\cal W$. Then, Equation~\ref{eq:flt} is the first law of thermodynamics (FLT)~\cite{Sekimoto1998, Seifert2012}. The only possible time reversal for passive dynamics is to choose ${\cal S}^+$ in Equation~\ref{eq:epr+}, so that ${\cal S}^+ = \dot{\cal Q}/T$. In this context, ${\cal S}^+$ coincides with the thermodynamic EPR due to contact of the system with the thermostat, which connects explicitly irreversibility, entropy production, and dissipation~\cite{Seifert2012}. In contrast, for active dynamics, the contribution $\bar{\cal  W}$ captures the energy cost to sustain the self-propulsion of particles~\cite{Pietzonka2019, Ekeh2020}, which is generally distinct from both $\cal W$ and $\cal Q$. When the potential is static ($\dot\alpha_n=0$), ${\cal S}^+ = \dot{\cal Q}/T$ still holds, yet ${\cal S}^+$ should not be interpreted as a thermodynamic EPR in general.

Although these considerations offer one consistent approach to the question of energy transfers for active particles, several other approaches are possible. Interestingly, one alternative definition of heat relies on replacing $\dot{\bf r}_i$ in Equation~\ref{eq:heat} by $\dot{\bf r}_i - \mu{\bf f}_i$. This amounts to regarding the self-propulsion as caused by a locally imposed flow, with particle displacements evaluated in the flow frame~\cite{Bo2019, Speck2018}. It yields vanishing heat in the absence of interactions, for the same reasons as led us to zero ${\cal S}^-$ in Equation~\ref{eq:epr+}. Also, some works addressing heat engines~\cite{Zakine2017, Holubec2020, Cates2021} have redefined heat by replacing $\sqrt{2\mu T}{\bm\xi}_i$ in Equation~\ref{eq:heat} with $\sqrt{2\mu T}{\bm\xi}_i + \mu{\bf f}_i$, thus considering the self-propulsion ${\bf f}_i$ as a noise with similar status to $\sqrt{2\mu T}{\bm\xi}_i$. This yields vanishing heat when the potential is static ($\dot\alpha_n=0$), by discarding all the energy dissipated by self-propulsion, allowing one to reinstate FLT, $U(t) - U(0) = {\cal W} - {\cal Q}$. Finally, the angular diffusion of ABPs is often itself regarded as thermal. This gives an angular contribution to $\cal Q$, proportional to $\int_0^t \sum_i \dot\theta_i(\dot\theta_i - \sqrt{2/\tau}\eta_i) dt'$, which vanishes for Equation~\ref{eq:abp}, but is generally non-zero when aligning interactions are present~\cite{Ekeh2020}. For AOUPs, interpreting the self-propulsion dynamics in terms of thermal damping and noise is less straightforward, although some studies have taken such a path~\cite{Mandal2017, Puglisi2017, Loos2020}.

Importantly, Equation~\ref{eq:dyn} does not resolve {\em how} particles convert fuel into motion. Accordingly, Equation~\ref{eq:heat} only captures the energy dissipated by the propulsion itself, ignoring contributions from underlying, {\it metabolic} degrees of freedom. Various schematic models describe the underlying chemical reactions in thermodynamically consistent terms, maintaining them out-of-equilibrium by holding constant the chemical potential difference between products and reactants~\cite{Speck2018, Seifert2018, Kapral2018}. For some of these models~\cite{Speck2018, Seifert2018}, the time evolution of ${\bf r}_i$ can be mapped into Equation~\ref{eq:dyn} when a chemical noise parameter is small~\cite{Pietzonka2019}. In this case, the difference between the partial and total heat, found by discarding or including reactions, is a constant, independent of the potential $U$. In a more refined model, the dynamics tracks the time evolution of chemical concentration, and the heat features explicitly the chemical current~\cite{Kapral2018}. Related features will emerge next, for continuum fields.

% -------------------------------------------------------------------------------

\subsubsection{Coarse-grained perspective: Energy transfers at hydrodynamic level}

For field theories, assuming that the free energy $\cal F$ depends on some set of control parameters $\alpha_n$, the work $\cal W$ is defined by analogy with Equation~\ref{eq:work} as
\begin{equation}\label{eq:work_hydro}
	{\cal W} = \int_0^t \sum_n \dot\alpha_n \frac{\partial\cal F}{\partial\alpha_n} dt' .
\end{equation}
This holds regardless of any additional nonequilibrium terms in Equations~\ref{eq:dyn_phi} and~\ref{eq:dyn_p}, and can be viewed as directly generalizing from the case of passive to active fields. In contrast, to define heat at a hydrodynamic level, one cannot depend {\it a priori} on a physical interpretation of the various dynamical terms as we made for particles. A minimal requirement, for emergence of FLT along the lines of Equation~\ref{eq:flt}, is that $\cal F$ in Equation~\ref{eq:work_hydro} is a genuine free energy, which stems from coarse-graining only the passive contributions of the microscopic dynamics. When theories are based solely on phenomenological arguments, this thermodynamic interpretation is absent.

In some cases, nonetheless, an active field theory ought to allow a thermodynamic interpretation, {\it e.g.} for the formation of membraneless organelles~\cite{Weber2019}. In contrast, we do not expect this if the same theory is used to describe the demixing of animal groups. To reinstate a thermodynamic framework, one can include {\it chemical fields} describing the fuel consumption underlying activity~\cite{Markovich2020}. This approach relies on linear irreversible thermodynamics (LIT)~\cite{Mazur}, as previously used to illuminate particle-based dynamics~\cite{Kapral2018}. LIT postulates linear relations between thermodynamic fluxes and forces, whose product determines the heat $\cal Q$. A class of {\it active gel} models were indeed first formulated in this way~\cite{Kruse2004, Prost2017}. In contrast, the field theories in Equations~\ref{eq:dyn_phi} and~\ref{eq:dyn_p} were not derived from LIT {\it a priori}, yet they can be embedded within it in a consistent manner.

To illustrate this, consider a scalar field $\phi$ obeying Equation~\ref{eq:dyn_phi}. The conservation law $\dot\phi = -\nabla\cdot{\bf J}$ relates this to the thermodynamic flux $\bf J$, whose conjugate force is $-\nabla(\delta{\cal F}/\delta\phi)$. Likewise a thermodynamic flux $\dot n$ describes some metabolic chemical process, with conjugate force a chemical potential difference $\Delta\mu$. Out-of-equilibrium dynamics is maintained by holding either $\dot n$ or $\Delta\mu$ constant~\cite{Ramaswamy2018}. Considering here the latter case, LIT requires that the active current ${\bf J}_\phi$ is linear in $\Delta\mu$ (although, like $\bf J$, it can be nonlinear in $\phi$), so that we identify ${\bf J}_\phi/\Delta\mu \equiv {\bf C}$  as an off-diagonal Onsager coefficient~\cite{Markovich2020}, yielding
\begin{equation}\label{eq:dyn_lit}
	\big[ {\bf J}, \dot n \big] = {\mathbb L} \bigg[ -\nabla\frac{\delta\cal F}{\delta\phi},
	\Delta\mu \bigg]
	+\hbox{\rm noise terms} ,
\end{equation}
with Onsager matrix obeying ${\mathbb L}_{\bf JJ} = \lambda{\bf I}$ (with ${\bf I}$ the d-dimensional identity);  ${\mathbb L}_{\dot n\dot n} = \gamma$, a {\it chemical mobility} (such that $\dot n = \gamma\Delta\mu$ in the absence of $\phi$ dynamics); and  ${\mathbb L}_{{\bf J}\dot n} = {\mathbb L}_{\dot n{\bf J}} = {\bf C}$. This last result encodes the famous Onsager symmetry which stems from the underlying reversibility of LIT~\cite{Onsager1931}, so that the form of the active current ${\bf J}_\phi$  also controls the $\phi$ coupling in the equation for $\dot n$. For ${\bf J}_\phi  = \Delta\mu \nabla((\nabla\phi)^2)$~\cite{Wittkowski2014}, then $\dot n = \gamma\Delta\mu-\nabla(\delta{\cal F}/\delta\phi)\cdot\nabla((\nabla\phi)^2)$. The covariance of the noise terms in Equation~\ref{eq:dyn_lit} is set directly by $\mathbb L$. The off-diagonal noise depends on $\phi$ through $\bf C$ and thus is multiplicative. The noise terms accordingly include so-called spurious drift contributions, which ensure that the LIT dynamics reaches Boltzmann equilibrium when neither $\dot n$ nor $\Delta\mu$ is held constant~\cite{Markovich2020}.

A crucial assumption of Equation~\ref{eq:dyn_lit} is that $\phi$ and $n$ are the {\it only} relevant hydrodynamic fields, so that, on the time and length scales at which they evolve, all other degrees of freedom are thermally equilibrated. Then, the heat $\cal Q$ is given in terms of the entropy production of these fields: ${\cal Q} = D \ln({\cal P}[{\bf J},\dot n]/{\cal P}^R[{\bf J},\dot n])$. The resulting {\em thermodynamic} EPR differs in form from the {\em informatic} EPR associated with the dynamics of $\phi$ alone (Equation~\ref{eq:epr}). Therefore, it cannot be equated with any of the various IEPRs for differently time-reversed pure $\phi$ dynamics given in Section~\ref{sec:path}. On the other hand, since Equation~\ref{eq:dyn_lit} follows LIT, the heat can be expressed in terms of currents and forces:
\begin{equation}\label{eq:heat_hydro}
	{\cal Q} = \int d{\bf r} \int_0^t dt' \bigg( {\bf J}\cdot\nabla\frac{\delta\cal F}{\delta\phi} + \dot n \Delta\mu \bigg) .
\end{equation}
Similar expressions arise for more elaborate field theories, such as polar fields~\cite{Markovich2020}. Note that $\cal Q$ is stochastic, as is $\cal W$ in Equation~\ref{eq:work_hydro}, even though both are defined at hydrodynamic level. Using the chain rule $\dot{\cal F} = \sum_n \dot\alpha_n (\partial{\cal F}/\partial\alpha_n)+ \int \dot\phi (\delta{\cal F}/\delta\phi) d{\bf r}$ and the conservation law $\dot\phi=-\nabla\cdot{\bf J}$, it follows that free energy $\cal F$, work $\cal W$, and heat $\cal Q$ are related as
\begin{equation}\label{eq:flt_field}
	{\cal F}(t) - {\cal F}(0) = {\cal W} - {\cal Q} + \int d{\bf r} \int_0^t dt' \dot n \Delta\mu .
\end{equation}
The energy balance in Equation~\ref{eq:flt_field} offers the hydrodynamic equivalent of the relation between the particle-based work $\cal W$, heat $\cal Q$, and energy $U$ in Equation~\ref{eq:flt}. When neither $\dot n$ nor $\Delta\mu$ is maintained constant, and $\Delta\mu$ derives from a free energy ($\Delta\mu=-\delta{\cal F}_{\rm ch}/\delta n$), the system achieves equilibrium, so that Equation~\ref{eq:flt_field} reduces to FLT with respect to the total free energy ${\cal F}+{\cal F}_{\rm ch}$. When the free-energy $\cal F$ does not change ($\dot\alpha_n=0$), the heat rate equals $\int d{\bf r}\,\langle \dot n\Delta\mu\rangle$, which illustrates that all the activity ultimately stems from the work done by chemical processes. At fixed $\Delta\mu$ (say), $\dot n$ depends on $\phi$, so that $\int d{\bf r}\,\langle \dot n\Delta\mu\rangle$ contains information similar to, but distinct from, Equation~\ref{eq:epr}, see discussion in Section~\ref{sec:coll}.

% -------------------------------------------------------------------------------

\subsection{Where and when activity matters}\label{sec:coll}

\subsubsection{Spatial and spectral decompositions of irreversibility}\label{sec:dec}

The various IEPRs introduced in Section~\ref{sec:epr} enable one to  delineate regimes where activity most affects the dynamics compared to equilibrium. Identifying such regimes can quantify the length and time scales where activity primarily matters, and sometimes pinpoint specific locations where self-propulsion plays an enhanced role. Within this perspective, the most appropriate choice of IEPR  is usually that which best reflects the dynamical symmetries of the emergent order (see {\bf Figure \ref{fig1}}). By eliminating gross contributions ({\it e.g.}, from propulsion pointing opposite to velocity) the right choice of IEPR can help identify dynamical features which break TRS more subtly, and help unravel how activity can create effects with no equilibrium counterpart.

\begin{figure}
	\includegraphics[width=.84\linewidth, trim=2.5cm 18.7cm 4cm 2.3cm, clip=true]{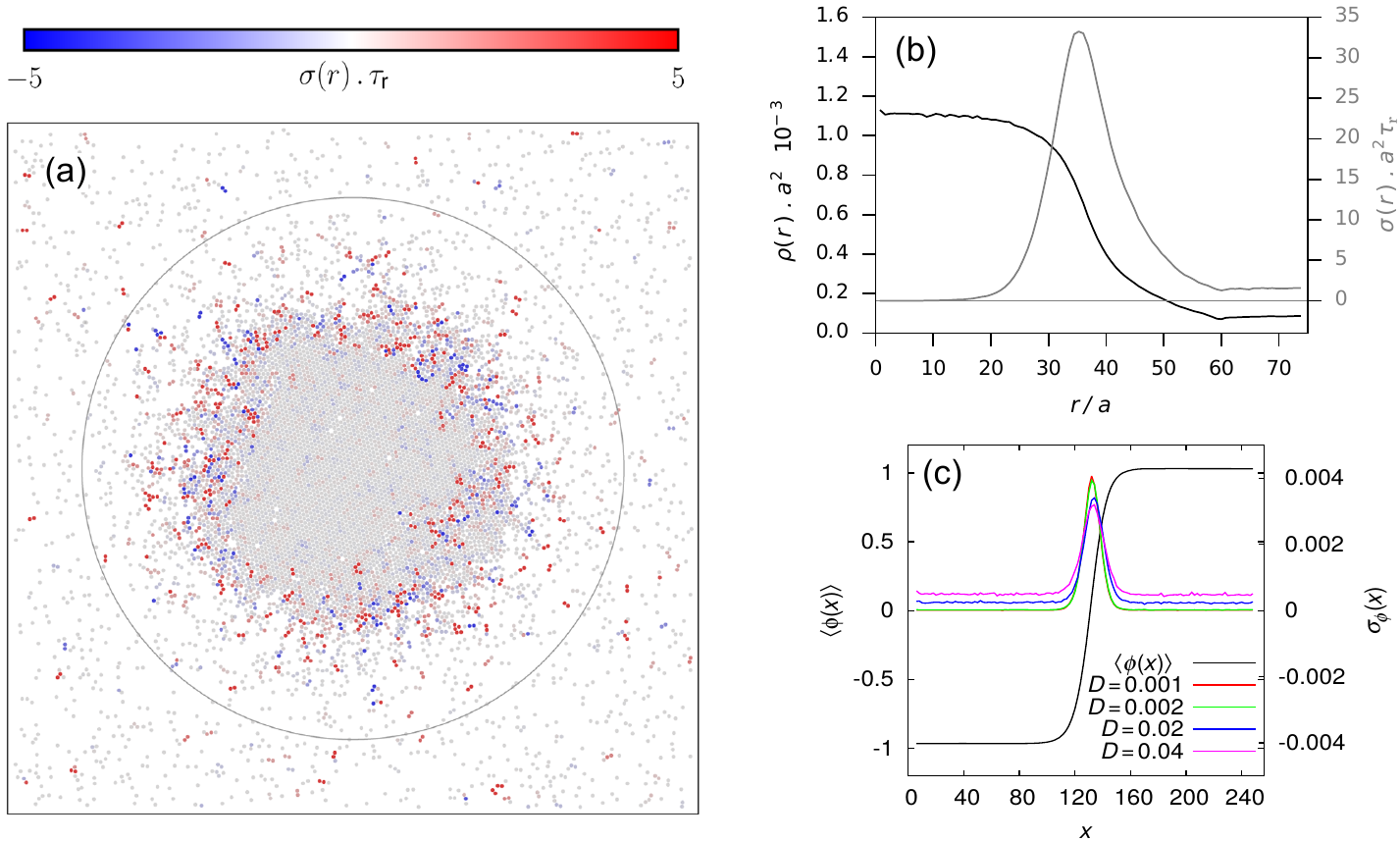}
	\caption{
		Spatial decomposition of the IEPR~\cite{Nardini2017, Martin2020}. (a) Phase separation of repulsive active particles, with individual $\sigma_i$ (color-coded {\it via} Equation~\ref{eq:epr_part}) enhanced at the liquid-vapor interface. (b) Time-averaged profiles of particle density $\rho$ and IEPR density $\sigma$ show that $\sigma$ is small in both bulks and peaked at the interface. (c) Similar behavior for a conserved active scalar field $\phi$.
	}
	\label{fig2}
\end{figure}

If thermal noise is omitted from Equation~\ref{eq:dyn}, only one IEPR can be defined. This is ${\cal S}_r$ (see Equation~\ref{eq:epr_r}), tracking positions only. For AOUPs with pairwise interaction $U=\sum_{i<j}V({\bf r}_i-{\bf r}_j)$, one finds a particle-based decomposition ${\cal S}_r = \sum_i \sigma_i$~\cite{Martin2020}, where
\begin{equation}\label{eq:epr_part}
	\sigma_i = \frac{\tau}{4\mu f_0^2} \sum_j \Big\langle \big[ (\dot{\bf r}_i-\dot{\bf r}_j) \cdot \nabla_i \big]^3 V({\bf r}_i-{\bf r}_j) \Big\rangle .
\end{equation}
This shows that particle $i$ contributes most to ${\cal S}_r$ when its neighbors $j$ have a large relative velocity $\dot{\bf r}_i-\dot{\bf r}_j$. Thus collisions between slow and fast particles are the main source of irreversibility. Interestingly, such collisions lie at the basis of the formation of particle clusters, which can potentially lead to nonequilibrium (or motility-induced) phase separation~\cite{Cates2015}. For a phase-separated density profile, Equation~\ref{eq:epr_part} allows one to distinguish the relative contributions from particles in different spatial zones. In the dilute phase, collisions are rare so that $\sigma_i$ stays small, whereas in a dense enough phase, particles barely move so that $\sigma_i$ is again modest. At interfaces, collisions between fast and slow particles, entering respectively from dilute and dense zones, lead to locally high IEPR; see {\bf Figure~\ref{fig2}}.

In the presence of thermal noise, one can also study ${\cal S}^\pm$~\cite{Shankar2018}. These IEPRs track the dynamics of both position and self-propulsion. The corresponding particle-based quantities, $\sigma_i^+=(1/T)\langle\dot{\bf r}_i\cdot{\bf f}_i\rangle$ and $\sigma_i^-=(\mu/T)\langle\nabla_iU\cdot{\bf f}_i\rangle$, which take ${\bf f}_i$ to be even and odd, are respectively minimal and maximal for particles at rest. This happens when $-\nabla_i U$ is equal and opposite to ${\bf f}_i$. In practice, $\sigma_i^+$ is low in the dense phase and high at the interface (like $\sigma_i$ above), whereas $\sigma_i^-$ decreases gradually from the bulk of the dense phase to the interface. Considering instead {\it polar} clusters, which emerge when aligning interactions are present, $\sigma_i^-$ is now low in the dense and dilute phases, with a higher value at the interface~\cite{Spinney2018}. Thus, for isotropic particles it is $\sigma_i^+$ that exposes the character of the irreversibility of self-propulsion at interfaces, whereas for aligning particles $\sigma_i^-$ does so. Each separately elucidates the role of interfacial self-propulsion to promote and stabilize clustering.

Turning now to field theories, we consider first the simplest scalar model of Equation~\ref{eq:dyn_phi}, with $\phi$ a particle density that is even under time reversal. In the spirit of Landau-Ginzburg theory, the nonequilibrium forcing term ${\bf J}_\phi$ is generally taken as a local function of $\phi$ and its gradients, and hence also even~\cite{Wittkowski2014, Tjhung2018}. The relevant measure of irreversibility reads ${\cal S}=\int \sigma({\bf r}) d{\bf r}$, where $\sigma = \langle {\bf J}\cdot{\bf J}_\phi \rangle/D$ can be identified as a local IEPR density~\cite{Nardini2017}. To lowest order, ${\bf J}_\phi$ contains three gradients and two fields, giving leading order (in noise) contributions where $\nabla\phi$ is large, as applies near the interface in a phase-separated profile (see {\bf Figure~\ref{fig2}}). This corroborates the particle-based results for active phase separation given above. (Similar principles govern various local IEPR densities in field theories of polar active matter~\cite{Borthne2020}.) In contrast to $\cal S$, the local chemical heat $\langle \dot n\Delta\mu\rangle$, found from Equation~\ref{eq:heat_hydro}, is high throughout the bulk phases with a (bimodal) dip across the interface~\cite{Markovich2020}. This reflects the fact that the underlying chemical reactions fruitlessly dissipate energy throughout uniform bulk phases (where $\bf C$ vanishes, modulo noise, so they barely affect the $\phi$ dynamics) whereas at interfaces such reactions locally do work, against the thermodynamic force $-\nabla(\delta{\cal F}/\delta\phi)$, to sustain the counter-diffusive current  ${\bf J}_\phi$.

Another challenge is to identify the time scales on which activity dominates over thermal effects. To partially quantify this, one can define a frequency-dependent energy scale $T_{\rm eff}(\omega) = \omega C_i(\omega)/(2 R_i(\omega))$. The autocorrelation of position reads $C_i(\omega) = \int e^{i\omega t} \langle {\bf r}_i(t)\cdot{\bf r}_i(0) \rangle dt$. The response $R_i$ for small perturbation of the potential ($U\to U - f \hat{\bf u}\cdot{\bf r}_i$, where $\hat{\bf u}$ is an arbitrary unit vector) reads $R_i(\omega) = \int \sin(\omega t) [\delta\langle {\bf r}_i(t)\cdot\hat{\bf u} \rangle/\delta f(0)]_{f=0} dt$. In the absence of self-propulsion,  $T_{\rm eff}$ reduces to the bath temperature $T$, restoring the FDT~\cite{Kubo1966}. The deviation $T_{\rm eff}-T$ at small $\omega$ identifies regimes where activity dominates, and has been measured in living systems~\cite{Guo2015, Ahmed2018, Gnesotto2018}. Interestingly, measuring the FDT violation offers a generic route to quantify irreversibility and dissipation. In passive systems, the Harada-Sasa relation~\cite{Sasa2005} explicitly connects the position autocorrelation $C_i$, the response function $R_i$, and the thermodynamic EPR, $\cal S$. For active particles, it can be generalized as~\cite{Nardini2016, Szamel2019}
\begin{equation}\label{eq:hs}
	{\cal S}^+ = \int \sum_i \frac{\omega}{\mu T} \Big[ \omega C_i(\omega) - 2 T R_i(\omega) \Big] \frac{d\omega}{2\pi} ,
	\quad
	{\cal S}_r = \int \sum_i \frac{\omega}{\mu} \Big[ 4\mu \omega \Gamma(\omega) C_i(\omega) - 2 R_i(\omega) \Big] \frac{d\omega}{2\pi} .
\end{equation}
The integrands in Equation~\ref{eq:hs} provides spectral decompositions of ${\cal S}^+$ and ${\cal S}_r$. Using models where ${\cal S}^+$ is directly proportional to heat, this decomposition has been measured experimentally to provide insights into the energetics of living systems~\cite{Toyabe2010, Ahmed2016}. Notably the integrand for the coordinate-only IEPR, ${\cal S}_r$, is no longer directly the FDT violation. Instead the irreversibility, when evaluated from fluctuations of position only, is quantified by the violation of a modified relation between correlation and response~\cite{Nardini2016}. For field theories, the FDT violation can again be expressed {\it via} correlation and response functions in the Fourier domain, depending on both temporal frequencies and spatial modes. It provides a spectral decomposition of IEPR which usefully extends the Harada-Sasa relation~\cite{Nardini2017, Murrell2021}. This helps identify the length and time scales primarily involve in the breakdown of TRS.  When considering theories with several fields, the quantification of irreversibility typically involves FDT violations associated with each one of them~\cite{Nardini2017, Ramaswamy2018}.

% -------------------------------------------------------------------------------

\subsubsection{Scaling of irreversibility with dynamical parameters}

For generic active dynamics, one can identify as dynamical parameters the coefficients of active terms of the dynamics. Examples include the strength of microscopic self-propulsion ${\bf f}_i$ for particle-based formulations as in Equation~\ref{eq:dyn}, and the coefficients of non-integrable terms in field theories, such as the leading-order contributions $\nabla((\nabla\phi)^2)$ and $(\nabla\phi)\nabla^2\phi$ within ${\bf J}_\phi$ in Equation~\ref{eq:dyn_phi}. Estimating how the various IEPRs depend on these activity parameters, and on temperature or density, allows one to delineate equilibrium-like regimes where TRS is restored either exactly, or asymptotically. Thus obtaining precise scalings for IEPRs is an important step towards understanding how far active dynamics deviates from equilibrium, {\em e.g.}, with a view to building a thermodynamic framework, starting with near-equilibrium cases.

Considering AOUPs in Equation~\ref{eq:aoup}, the persistence time is a natural parameter that controls the distance from the equilibrium limit at $\tau = 0$. This approach allows a systematic perturbative expansion in $\tau$ (at fixed $T_a$) for the steady state~\cite{Nardini2016, Martin2020, Martin2020b}. Expanding the IEPR (here ${\cal S}_r$) shows the irreversibility to vanish at linear order in $\tau$, even as the statistics differ from the Boltzmann distribution $\sim e^{-U/T_a}$~\cite{Nardini2016}. In other words, there exists a regime of small persistence where the steady state is distinct from that of an equilibrium system at temperature $T_a$, yet TRS is restored asymptotically. The existence of such a regime lays the groundwork for extending standard relations of equilibrium thermodynamics.

Various works have studied the behavior of ${\cal S}_r$ beyond the small $\tau$ regime. They have shed light on a non-monotonicity of ${\cal S}_r(\tau)$ for dense systems~\cite{Szamel2020b}, and also for particles confined in an external potential~\cite{Bo2020}. Note that, in the latter case,  the dependence of ${\cal S}_r$ on bath temperature $T$ is also non-monotonic in general~\cite{Martin2020b}. These results suggest that departures from equilibrium can decrease with persistence, and increase with thermal fluctuations. Note however that the IEPR per particle has units of inverse time, see Equation~\ref{eq:epr}, so that a saturating entropy production per persistence time gives decreasing ${\cal S}_r(\tau)$. Turning to repulsive particles, the IEPRs ${\cal S}^+$ and ${\cal S}^-$ respectively decrease and increase with $\tau$, up to the onset of MIPS~\cite{Spinney2018}. This is consistent with their being respectively low and high in clustered regions, and is unaffected by the rescaling ${\cal S}^\pm \to{\cal S}^\pm\tau$~\cite{Suri2020}.

Another parameter controlling nonequilibrium effects is the density $\rho$. In homogeneous state, since ${\cal S}^-$ increases as the dynamics slows down, it increases (decreases) with $\rho$ for isotropic (aligning) particles. In practice, some detailed scalings can be obtained for isotopic pairwise interactions of the form $U = \sum_{i<j}V({\bf r}_i-{\bf r}_j)$. For weak interactions, namely when the amplitude of $V$ is small compared to those of thermal noise and self-propulsion, ${\cal S}^-$ becomes linear in $\rho$~\cite{Suri2020}. In general, it can be written in terms of density correlations:
\begin{equation}\label{eq:epr_struc}
	\frac{T{\cal S}^-}{\mu} = \rho \int \big[ (\nabla V)^2 - T\nabla^2 V \big] g_2({\bf r}) d{\bf r} + \rho^2 \iint (\nabla V({\bf r}))\cdot(\nabla V({\bf r'})) g_3({\bf r},{\bf r}') d{\bf r} d{\bf r}' ,
\end{equation}
where $g_2 = \sum_{i\neq j}\langle\delta({\bf r}-{\bf r}_i+{\bf r}_j)\rangle/(N\rho)$ and $g_3 = \sum_{i\neq j\neq k}\langle\delta({\bf r}-{\bf r}_i+{\bf r}_j)\delta({\bf r}'-{\bf r}_i+{\bf r}_k)\rangle/(N\rho)^2$ are two-point and three-point density correlators~\cite{Suri2019}. The integrand in Equation~\ref{eq:epr_struc} can be regarded as an integral form of the Yvon-Born-Green relation, which constrains $g_2$ and $g_3$ for passive particles~\cite{Hansen}, so that the violation of this relation provides access to ${\cal S}^-$ for active ones. A related form of Equation~\ref{eq:epr_struc} has been proposed in terms of $g_2-g_{2,\rm eq}$, where $g_{2,\rm eq}$ is the two-point correlator for passive particles with the same potential $U$~\cite{Suri2019, Suri2020b}.

In field theories, IEPRs are generically either linear or quadratic in the activity parameters (for small parameters) depending on whether they break symmetries of the passive theory~\cite{Nardini2017}. The IEPR scalings with $D$ are mechanistically revealing both for scalar~\cite{Nardini2017} and polar~\cite{Borthne2020} models. As previously noted, the spatial dependence of IEPR density is informative: independent scalings can be seen in different bulk phases, and at their interfaces. In general one finds the following leading order behavior. (i)~$\sigma\sim D^{-1}$: the deterministic part of the dynamics already breaks TRS, and very unlikely noise realizations are needed to recreate, for the reversed paths, the opposite of a deterministic forward motion. (ii)~$\sigma\sim D^{0}$: TRS is unbroken deterministically but violated at lowest order in fluctuations. This may require some spatial symmetry breaking at deterministic level, so that the background fields act as a ratchet or rectifier for the fluctuations. (iii)~TRS is broken only at higher order, which gives $\sigma\sim D^{1}$ in all cases so far studied, although higher powers are not ruled out.

For scalar fields exhibiting bulk phase separation, the deterministic dynamics give a stationary mean-field profile, so that there is no $D^{-1}$ contribution. TRS is broken at leading order in fluctuations because of the interface between phases ({\bf Figure \ref{fig2}}). In uniform bulk phases, irreversibility emerges only at next order, $\sigma\sim D^{1}$, suggesting that it arises from interactions between fluctuations. Note that active scalar models can also predict entirely new phases, whose $\sigma$-scalings remain subject to investigation~\cite{Tjhung2018}. The behavior of the thermodynamic EPR, $\langle \dot n\Delta\mu\rangle({\bf r})/D$, found by embedding the same model within LIT and including chemical processes, is quite different~\cite{Markovich2020}. This scales as $D^{-1}$, but with a reduced amplitude in interfacial regions (see Section \ref{sec:dec}). Moreover, for one scalar model with both conserved and nonconserved dynamics, TRS violations are gross ($D^{-1}$) if these two contributions to $\dot\phi$ are separately monitored, but perfect reversibility holds if not~\cite{Li2021}.

For dynamics with a scalar density field (even under time reversal) and a polarization (even or odd)~\cite{Borthne2020, Ramaswamy2018}, the IEPRs scalings depend on whether the density current ${\bf J}$ is retained or ignored (see Equations~\ref{eq:dyn_phi} and~\ref{eq:dyn_p}, and Section~\ref{sec:path}); we assume the latter here. There remain two IEPRs, ${\cal S}^{\pm}$ according to whether ${\bf p}$ is even ($+$) or odd ($-$). In the phase comprising traveling bands or clusters (polarized high density regions propagate along ${\bf p}$), both of ${\cal S}^\pm$ scale as $D^{-1}$, since such deterministic dynamics breaks TRS. Moreover, because the density wave has a steep front and a shallow back, flipping ${\bf p}$ does not restore deterministic-level TRS ({\bf Figure \ref{fig1}}). In the phase of uniform $\langle {\bf p}\rangle$, one finds instead ${\cal S}^\pm \sim D^0$; similar arguments might apply now to fluctuations instead of deterministic motion.

As emphasized already, IEPRs give varying information depending on which degrees of freedom are retained, and which ignored. In particular, the universal properties of the critical point for active phase separation (including MIPS) can be studied by progressive elimination of degrees of freedom {\it via} the Renormalization Group, which could potentially be expected to give rise to {\it emergent reversibility}. Remarkably, Ref.~\cite{Caballero2020} has established that (i)~the active critical point is in the same universality class as passive phase separation with TRS, but (ii)~there is no emergent reversibility, in the sense that the IEPR {\em per spacetime correlation volume} does not scale towards zero at criticality (it remains constant above 4 dimensions, and it can even diverge below). This scenario, referred to as {\it stealth entropy production}, can be rationalized by arguing that the scaling of IEPR has a non-trivial critical exponent in the universality class shared by active and passive systems. Because no coarse-graining can make a passive system become active, IEPR has zero amplitude in the passive cases previously thought to define the universality class.

% ===============================================================================

\section{BIASED ENSEMBLES OF TRAJECTORIES}\label{sec:bias}

\subsection{Dynamical bias and optimal control}\label{sec:traj-thermo}

Biased ensembles generalize the canonical ensemble of equilibrium statistical mechanics from microstates to trajectories~\cite{Maes1999gibbs, Lecomte2007, Garrahan2009}. They have proven useful in simple models of interacting particles~\cite{Bodineau2004,Appert2008,Jack2015,Dolezal2019} as well as glassy systems~\cite{Garrahan2007, Hedges2009} and beyond~\cite{GarrahanLesanovsky2010,Weber2014,Nemoto2019}. They are constructed by biasing the value of a physical observable which we denote here by ${\cal B}$. The choice of this observable depends on the physical system of interest, {\it e.g.} the heat dissipated during a dynamical trajectory, or the displacement of a tagged particle.

Let $X$ denote a dynamical trajectory of an active system, over a time period $[0,\tobs]$.  Following Section~\ref{sec:path}, the probability of this trajectory can be represented as ${\cal P}[X]=p_0[X] {\rm e}^{-{\cal A}[X]}$ where ${\cal A}$ is the action and $p_0$ is the probability of the initial condition. A biased ensemble is defined by a probability distribution over these trajectories:
\begin{equation}
	{\cal P}_s[X]= \frac{1}{{\cal Z}(s,\tobs)} p_0[X] \exp\left( -{\cal A}[X] - s{\cal B}[X] \right) ,
\label{equ:Ps}
\end{equation}
where $s$ is the biasing field and ${\cal Z}(s,\tobs) = \langle {\rm e}^{-s{\cal B}} \rangle$ for normalisation. Comparing with the canonical ensemble, one may identify ${\cal B}$ and $s$ as an (extensive) physical observable and its conjugate (intensive) field, respectively. The field $s$, according to its sign, biases the distribution towards higher or lower ${\cal B}$. Importantly, the bias has no prejudice about the dynamical mechanism by which ${\cal B}$ changes: this allows the system to access fluctuation mechanisms for this quantity that might not have been anticipated a priori~\cite{Hedges2009, Weber2014, Nemoto2019}. Consistent with this observation, Equation~\ref{equ:Ps} can also be derived by a maximum entropy computation for trajectories with non-typical $\cal B$~\cite{Simha2008,Dill2013}.

The next step is to take a limit of large time $\tobs$, analogous to the thermodynamic limit of large system size in equilibrium statistical mechanics.  Since ${\cal Z}(s)$ is analogous to a partition function, one may define a dynamical free energy density as
\begin{equation}\label{eq:psi}
	\psi(s) = \lim_{\tobs\to\infty} \frac{1}{\tobs} \log {\cal Z}(s,\tobs) .
\end{equation}
This construction is natural if the observable ${\cal B}$ scales extensively with the time $\tobs$, which will be assumed below; $\psi$ is then a scaled cumulant generating function for ${\cal B}$~\cite{Touchette2009}. In contrast to the standard ensembles of equilibrium statistical mechanics, Equation~\ref{equ:Ps} does not describe practical experimental systems. However, there are several theoretical contexts in which such distributions are relevant. Examples include fluctuation theorems, in which ${\cal B}$ is a measure of irreversibility, whose probability distributions have symmetries related to TRS, yielding $\psi(s)=\psi(1-s)$~\cite{Lebowitz1999, Maes1999gibbs, Seifert2012}. Biased ensembles are also deeply connected with large deviation theory~\cite{denH-book}, applied to time-averaged quantities~\cite{Touchette2009,Jack2020}, which describes rare events where ${\cal B}/\tobs$ differs significantly from its typical value. Viewing $\psi(s)$ as a dynamical free energy, its singularities can be interpreted as dynamical phase transitions. In most cases, such transitions require a limit where both $\tobs$ and the system size go to infinity.

Given that biased ensembles do not mimic typical experimental situations, a natural question is whether the response to the bias $s$ can be connected with its response to some physical perturbation.  Such a connection is provided by optimal control theory~\cite{bertsekas-book}, which is also intrinsically related to large-deviation theory~\cite{dupuis-book,Chetrite2015var,Jack2015b}. In fact, biased ensembles show how ${\cal B}$ can be reduced or increased, and the corresponding mechanisms are as close as possible to the original dynamics. We now discuss a method for identifying a physical system that mimics a given biased ensemble~\cite{dupuis-book,Chetrite2015,Chetrite2015var,Jack2020}. To reduce the level of technicality, we exclude transient behavior which happens for times close to the beginning or end of a trajectory. Biased ensembles can then be accurately reproduced by the {\it optimally-controlled system} (OCS), also called auxiliary process~\cite{Jack2010} or driven process~\cite{Chetrite2015}. To find the OCS, consider a wider class of controlled systems, obtained by modifying the equations of motion of the original active model. For any controlled system, the Kullback-Leibler (KL) divergence measures how its path probability ${\cal P}^{\rm con}[X]= p_0[X] {\rm e}^{-{\cal A}^{\rm con}[X]}$ differs from ${\cal P}_s$:
\begin{equation}
	D_{\rm KL}\left( {\cal P}^{\rm con} || {\cal P}_s \right) = \int  {\cal P}^{\rm con}[X] 
\log \frac{ {\cal P}^{\rm con}[X] }{ {\cal P}_s[X] } {\cal D}X ,
\label{equ:KL-con}
\end{equation}
where ${\cal D}X$ denotes a path integral. The OCS has $\lim_{\tobs\to\infty} \tobs^{-1} D_{\rm KL}\left( {\cal P}^{\rm con} || {\cal P}_s \right) =0$, so its differences from ${\cal P}_s$ are small. From Equations~\ref{equ:Ps},~\ref{eq:psi} and~\ref{equ:KL-con}, one can deduce~\cite{Jack2020}
\begin{equation}
	\lim_{\tobs\to\infty} \frac{1}{\tobs} \big\langle {\cal A}^{\rm con}[X] - {\cal A}[X] - s {\cal B}[X] \big\rangle_{\rm con} \leq \psi(s) ,
\label{equ:bound-psi}
\end{equation}
where the average $\langle\cdot\rangle_{\rm con}$ is computed for the controlled system.  Equation~\ref{equ:bound-psi} provides a lower bound on $\psi$, which becomes an equality for the OCS.  Among all the processes with any given $\langle {\cal B}\rangle_{\rm con}$, the OCS minimizes $ \left\langle {\cal A} - {\cal A}^{\rm con}\right\rangle_{\rm con}$. Hence, it provides a mechanism for achieving an atypical value of $\cal B$, while remaining as close as possible to the original dynamics.

Knowing ${\cal A}$, ${\cal A}^{\rm con}$, and ${\cal B}$, it suffices to maximize the bound in Equation~\ref{equ:bound-psi} over the controlled system to obtain the OCS. The details of this computation depend on the unbiased dynamics of interest, as discussed extensively in~\cite{Jack2020,dupuis-book,Chetrite2015var,Chetrite2015}. In practice, the OCS usually differs from the original system in two ways: (i)~the forces in the original model get modified by terms that depend in simple and explicit ways on $s$ and ${\cal B}$, and (ii)~other control forces are added as gradients of an optimal control potential $U^{\rm opt}$, which may be computed by solving either an eigenvalue problem (following Doob~\cite{Chetrite2015}) or a non-linear partial differential equation~\cite{bertsekas-book}. We emphasize that the optimal control forces do not generically appear as {\em external forces} in the dynamics, they can correspond to complicated interactions which may be long-ranged. Although exact results for $U^{\rm opt}$ are rare~\cite{Touchette2016,Mallmin2020,Keta2020}, Equation~\ref{equ:bound-psi} is still useful. For example, using a controlled system with small KL divergence dramatically speeds up the numerical sampling of biased ensembles~\cite{Nemoto2016,Nemoto2017first,Limmer2018, Whitelam2019}. Also, minimization of the KL divergence over simple classes of controlled system provides useful insight into relevant fluctuation mechanisms, even without an exact solution.

% -------------------------------------------------------------------------------

\subsection{Phase transitions and symmetry breaking}\label{sec:biased-PT}

\subsubsection{Biased ensembles for active Brownian particles}

\begin{figure}
	\includegraphics[width=88mm]{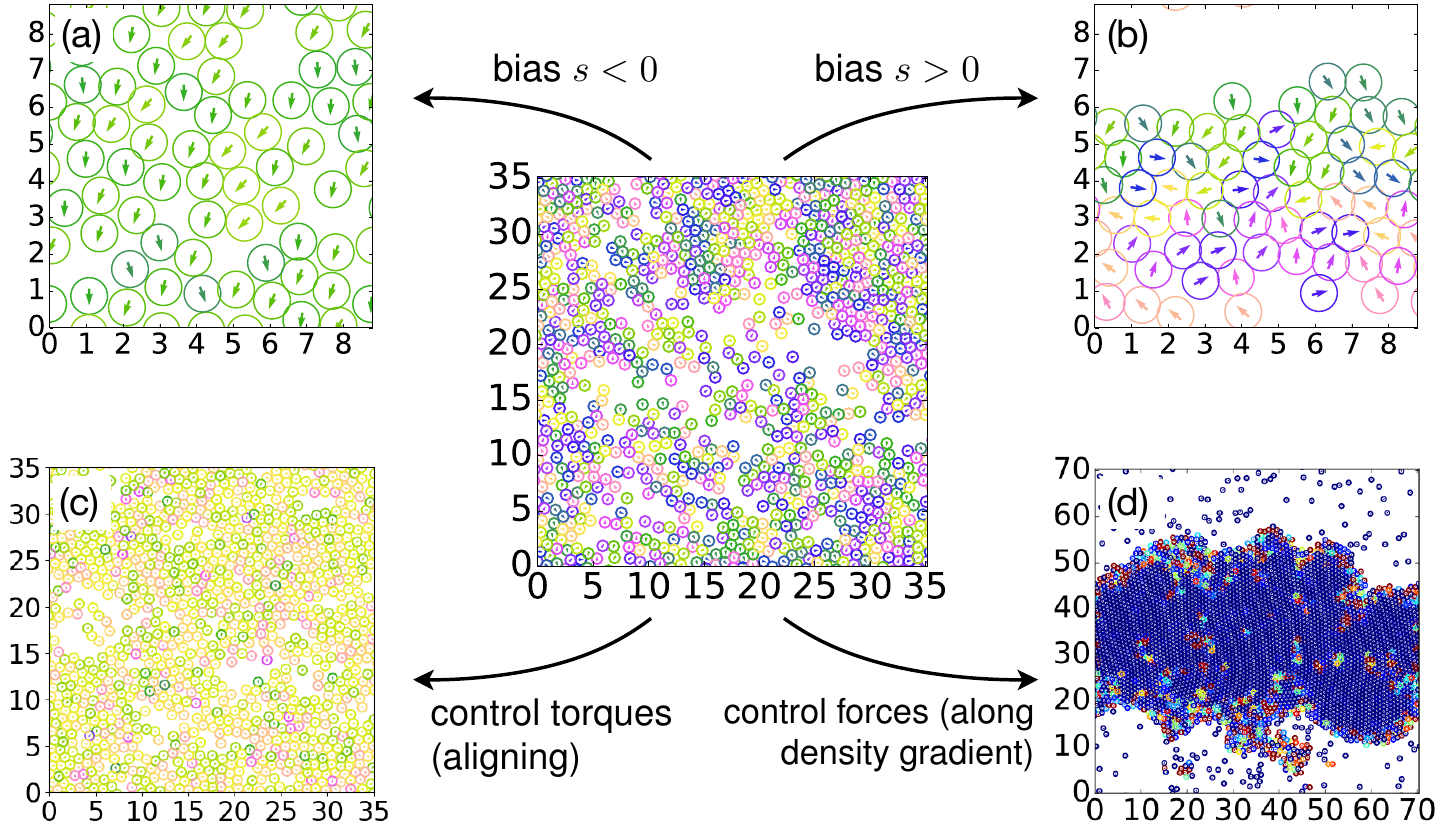}
	\caption{
	Behavior in biased ensembles of ABPs, and comparison with controlled systems~\cite{Nemoto2019,Keta2020}. The central panel shows a snapshot of the unbiased dynamics ($\bar{\cal  W}\simeq\langle \bar{\cal  W}\rangle$). Panels (a,b) show similar shapshots of collective motion and PSA states (respectively biased ensembles with $s < 0$ and $s > 0$) associated with atypical $\bar{\cal  W}$. Particles are colored according to their orientations. Panels (c,d) show how biased ensembles  can be mimicked by control forces (or torques). In (c), an infinite-ranged interaction favors orientational alignment; particles with similar colors are aligned with each other, as in (a). In (d), control torques tend to align orientations along density gradients. Particles are colored according to their control torque strength, which tends to be largest for boundary particles, facilitating phase separation.
	}
	\label{fig3}
\end{figure}

To illustrate the behavior of biased ensembles, we discuss a guiding example for a system of ABPs~\cite{Nemoto2019,Keta2020}. Section~\ref{sec:bias-other} below presents biased ensembles of active systems in a wider context. {\bf Figure~\ref{fig3}} displays the behavior of biased ensembles for a two-dimensional system of ABPs. Trajectories are biased according to their active work, that is ${\cal B}=\bar{\cal  W}$ in Equation~\ref{equ:Ps}, with
\begin{equation}
	{\bar{\cal  W}}[X] = \sum_i {\bar{\cal  W}}_i[X], \qquad {\bar{\cal  W}}_i[X] =  \frac{1}{w_0}  \int_0^\tobs  \dot{\bf r}_i \cdot {\bf f}_i dt' ,
\end{equation}
where the normalization $w_0$ is chosen so that $\langle {\bar{\cal  W}}_i \rangle = \tobs$ for non-interacting (or sufficiently dilute) particles. Physically, $\bar{\cal  W}$  measures how effectively the particles' propulsive forces are converted into motion: freely swimming particles have $({\bar{\cal  W}}_i/\tobs)\approx 1$ while those in crowded environments are impeded by their neighbors, resulting in $({\bar{\cal  W}}_i/\tobs)\approx 0$. Negative ${\bar{\cal  W}}_i$ means that a particle's motion is opposite to its self-propulsive force, which occurs rarely. Since $\bar{\cal  W}$ is closely related to IEPR (Section~\ref{sec:path}), it follows a fluctuation theorem~\cite{Seifert2012}: the dynamical free energy of the biased ensemble obeys $\psi(w_0-s)=\psi(s)$.

{\bf Figure~\ref{fig3}} shows how the biasing field modifies the behavior of this system. The following discussion applies for $s<w_0/2$, the behavior for larger $s$ can then be obtained {\it via} the fluctuation theorem. For $s>0$, the system enters a state that is phase-separated and dynamically arrested (denoted PSA). Consistent with the arguments of Section~\ref{sec:traj-thermo}, this happens because phase separation is the most natural (or least unlikely) mechanism for particles to collectively reduce their active work, {\it via} local crowding effects. Alternative mechanisms do exist for reduced active work, for example dilute ABPs would have ${\bar{\cal  W}}\approx 0$ during rare fluctuations where the noise term $\sqrt{2\mu T}\bm{\xi}_i$ in Equation~\ref{eq:dyn} is opposite to the propulsion force ${\bf f}_i$. Collectively, though, these trajectories have much larger action ${\cal A}$ than the PSA state; and for a given value of $\bar{\cal  W}$, the biased ensemble is dominated by the trajectories of least action. Hence, one observes PSA in the biased ensemble.

This behavior is closely related to biased ensembles for equilibrium systems, which can be analyzed at field-theoretic (or fluctuating hydrodynamic) level, using macroscopic fluctuation theory (MFT)~\cite{Bertini2015}. The particle density is a locally conserved field, so large-scale density fluctuations relax on a time scale proportional to $L^2$ (where $L$ is the system size). For large systems, this slow time scale decouples from particles' rapid microscopic motion, leading to a theory for the density alone. Construction of this theory involves scaling the space-time co-ordinates, so that the biasing field $s$ enters {\it via} a scaling variable $\lambda = sL^2$. For equilibrium steady states, MFT predicts dynamical phase transitions at finite $\lambda$, corresponding to a bias $s$ of order $L^{-2}$~\cite{Appert2008}. This argument applies also in the active case: the result is that sufficiently large systems should phase separate as soon as $s$ is positive.

To understand this result physically, recall the connection of biased ensembles to optimal control theory (Equation~\ref{equ:bound-psi}). As well as biased systems, {\bf Figure~\ref{fig3}} shows those to which control forces have been applied. To mimic PSA states, a suitable control strategy is to create a macroscopic cluster, whose boundary particles have their self-propulsion forces pointing inwards (similar to MIPS states). This can be achieved by applying torques to the ABP orientations, so that they point along the local density gradient, stabilizing the phase-separated cluster, and increasing its density towards dynamical arrest~\cite{Nemoto2019}. Such control forces (or torques) act mostly on the boundary particles, so that the difference between the action terms in Equation~\ref{equ:bound-psi} scales as $L$ ({\em i.e.}, sub-extensive in the volume $L^2$ of the system). This is consistent with the irreversibility measure $(1/T)\langle\dot{\bf r}_i\cdot{\bf f}_i\rangle$ being high at the interface for a phase-separated profile in the unbiased dynamics (Section~\ref{sec:coll}). Since the subextensive control forces lead to an extensive change in $\bar{\cal  W}$, one can confirm that PSA is expected for all $s>0$~\cite{Nemoto2019}. Note that control using torques is efficient for ABPs, since it harnesses the inherent self-propulsion to stabilize the cluster. Alternatively, applying control forces to particle positions, as in~\cite{Whitelam2018}, would result in a larger (though still subextensive) action.

Returning to Fig.~\ref{fig3}, one sees for $s<0$ that the particles align their orientations to create a state of collective motion. Since ABPs move with fixed speed, this alignment suppresses particle collisions, which is a natural way to increase the active work $\bar{\cal  W}$.  Spontaneous breaking of rotational symmetry may be surprising, since aligning interactions are completely absent from the ABP model.  However, numerical results~\cite{Nemoto2019,Keta2020} show that this biased ensemble can be described rather accurately by a controlled model with mean-field (infinite-ranged) interactions among the ABP orientations.

It is clear that any field-level description of collective motion requires a polarization field, in order to capture such alignment. Hence, collective motion cannot be described by a fluctuating hydrodynamic theory for the density alone, contrary to the PSA state.   Also, the collective motion phase is found for biasing fields beyond a threshold $s^* \simeq -1/\tau$, rather than at infinitesimal $s$ as for PSA. These observations are related: in the PSA state, the response to the bias takes place by a slow (hydrodynamic) field, for which the natural scale is $s\sim L^{-2}$; for the collective motion state, the response occurs by a fast field, which requires $s$ of order unity.  In fact, responses to small biasing fields $s$ are generically dominated by slow hydrodynamic modes~\cite{Jack2020}. If the quantity ${\cal B}$ couples to these, it responds strongly at small $s$~\cite{Bodineau2004,Appert2008}. On the other hand, for $s$ of order unity, an accurate description of the large-scale, bias-induced atypical dynamics may require analysis of fields (like polarization in this example) that are negligible for hydrodynamic descriptions of typical trajectories.

We now consider a coarse-grained (hydrodynamic) theory that captures the collective motion in biased ensembles with low active work. This reinforces the correspondence between particle models and field-theoretic descriptions. As noted above, describing the onset of collective motion at $s\sim s^*$ requires consideration of a polarization field $\bP$ alongside the density $\phi$, as in Equations~\ref{eq:dyn_phi} and \ref{eq:dyn_p}:
\begin{equation}
	\bJ = v(\phi) \bP -  D_{\rm c}(\phi) \nabla \phi+ \sqrt{2M(\phi)} \bm{\Lambda}_\phi ,
	\qquad
	\dot \bP =  \bF(\phi,\bP) + \sqrt{2D} \bm{\Lambda}_p ,
\end{equation}
where $v$ is self-propulsion, $D_{\rm c}$ is a (collective) diffusion constant, $M$ a mobility, $\bF$ an effective thermodynamic force~\cite{Keta2020}. To analyze biased ensembles, we need the dependence of the active work on $\phi,\bP$. At hydrodynamic level, it is consistent to approximate this as an integral of a local work rate $w$ as ${\bar{\cal  W}} \approx \int_0^t \int w(\phi,\bP) \, {d}{\bf r} \, {d}t' $. Here, $w$ is averaged over a mesoscopic region and a time interval where density and polarization have prescribed values.

To understand collective motion, the key point is that the active work tends to be reduced by alignment (because collisions are reduced); hence for small polarization $\bP$: $w(\phi,\bP) \simeq w(\phi,\bm{0}) - c_w(\phi) |\bP|^2$, where $c_w>0$ is the coupling between active work and polarization. Construction of the coarse-grained action for this minimal theory confirms that $\bP$ is a fast field, while $\phi$ is slow (hydrodynamic). It is consistent to assume that $\bF(\phi,\bP)\simeq -c_{\rm F}(\phi) \bP$ where $c_{\rm F}>0$ is the strength of the restoring force towards $\bP=\bm{0}$.  Hence, for trajectories $X$ with homogeneous density and small (constant) polarization $\bP$, one has ${\cal P}_s[X]\sim {\rm e}^{-{\cal A}_{\rm pol}[X]}$ (see Equation~\ref{equ:Ps}), where ${\cal A}_{\rm pol} \propto \left( \frac{ c_{\rm F}^2 }{4} + s c_w \right) |\bP|^2 + O(|\bP|^4)$. Minimizing the Landau-like action ${\cal A}_{\rm pol}$ reveals that spontaneous symmetry breaking occurs for $s<s_c$, where $s_c = -c_{\rm F}^2/(4c_w) < 0$. This connects to particle-level results {\it via} the scaling estimates $c_w\sim 1$ and $\gamma c_{\rm F}^2 \sim \tau^{-1}$, yielding symmetry breaking at $s_c \sim -1/\tau$~\cite{Keta2020}.

% -------------------------------------------------------------------------------

\subsubsection{Generic features of biased ensembles in active systems}\label{sec:bias-other}

We now discuss the insights from the previous example, in a broader context. First, note that biasing the active work can create both orientational order and density modulations; this stems from the couplings between fluctuations of the active work and those of local alignment and density. The first of these can already be deduced {\it via} exact biased-ensemble results in a system with just two active particles~\cite{Mallmin2020,Keta2020}: pairs with parallel alignment tend to have fewer collisions, so these states are promoted when biasing to higher active work (or propulsion efficiency). Broader insight stems from the optimal control interpretation of biased ensembles (Section~\ref{sec:traj-thermo}). For ABPs biased by the active work, the collective motion at large $\bar{\cal W}$ shows that local alignment is an effective route towards efficient swimming~\cite{Nemoto2019,Mallmin2020}, suggesting a design strategy to achieve active particles with desired properties.

Similar arguments explain the coupling of active work to local density and hence to phase separation, because crowding reduces the particles' swimming speed, as found by fluctuating hydrodynamic arguments~\cite{Nemoto2019,GrandPre2020,Keta2020}, perturbative calculations~\cite{Suri2019, Suri2020} and variational computations~\cite{Whitelam2018,GrandPre2020}.  Many other active systems couple alignment, density, and work by similar mechanisms; so this behavior under bias should be generic. Indeed, the results of~\cite{Keta2020,Mallmin2020,Nemoto2019,Suri2019, Suri2020,Whitelam2018} cover a range of different active systems, including lattice models as well as AOUPs and ABPs. Dynamical arrest in biased ensembles also has a counterpart in passive glassy systems~\cite{Garrahan2007,Hedges2009, Jack2020}, and can be explained by general principles that couple biasing fields to metastable states~\cite{Jack2020}.

All these results concern systems where the observable $\cal B$ is defined as a sum over all particles. Instead, $\cal B$ may refer to a tagged particle within a large system~\cite{Cagnetta2017, GrandPre2018, Cagnetta2020}. Perhaps surprisingly, results for equilibrium systems show that applying such bias to a single particle can generate a macroscopic response~\cite{Dolezal2020}. This is because of coupling of the biased quantity to the slow (hydrodynamic) density fluctuations, as described by MFT. In active systems, the results of~\cite{Cagnetta2017,Cagnetta2020} indicate a rich behavior for large deviations of the single-particle active work, which appears related to a coupling with collective (long-ranged) density fluctuations. On the other hand, fluctuations of the single-particle current were explained in~\cite{GrandPre2018} by a local argument based on a density-dependent velocity.

% ===============================================================================

\section{CONCLUSION}\label{sec:conc}

The tools of stochastic thermodynamics, first introduced for thermodynamically consistent models~\cite{Sekimoto1998, Seifert2012, Lebowitz1999}, provide fruitful insights when analyzing the consequences of irreversibility in active dynamics. The framework has to be carefully adapted, since various thermodynamic relations do not carry over {\it a priori}. For instance, the usual connection between irreversibility, entropy production and dissipation of energy needs to be revisited (Sections~\ref{sec:path} and~\ref{sec:ener}). The first law of thermodynamics no longer holds in its standard form, a fact that strongly influences the behavior of engines using an active working substance~\cite{Cates2021}. However, the informatic entropy production (IEPR) can be defined without reference to the first law, directly in terms of forward and backward path probabilities. This makes it a useful measure of irreversibility in active systems. In contrast with equilibrium, care is now needed in choosing which coarse-grained quantities change sign on time reversal: different choices yield different information on how time-reversal symmetry is broken.

Importantly, measuring irreversibility {\it via} IEPR can provide a systematic approach to evaluate the distance of active dynamics from an equilibrium reference. This allows one to pinpoint regimes (such as specific bulk phases, or interfaces between these) where activity plays a major role, and rationalize the mechanisms whereby self-propulsion at the particle scale stabilizes collective effects with no equilibrium equivalent (Section~\ref{sec:coll}). As reviewed above, this approach has been validated in several settings, involving either particle-based or field-level dynamics for scalar and polar materials, but much remains to be explored beyond these examples. In particular, it would be interesting to consider active nematics~\cite{Chate2020}.

Very recent progress in stochastic thermodynamics has opened new avenues whose implications for active matter remain largely unexplored. In particular, thermodynamic uncertainty relations (TURs) place bounds on irreversibility in the presence of currents~\cite{Horowitz2020}. For active systems, it is not yet known how different types of emergent order affect the tightness of these bounds. Understanding this could allow delineation of regimes where TURs are most accurate, with direct relevance for experimental studies of living systems.

Finally, biased ensembles (Section~\ref{sec:bias}) offer promise for the reverse engineering of active materials~\cite{Suri2019, Nemoto2019, Mallmin2020}. Within this perspective, one pre-defines a desired macroscopic property in terms of a specific biasing observable. For a given starting material, the sampling of biased trajectories determines a strategy to tune microscopic interactions and achieve the target property. This approach is potentially also relevant when developing strategies to control the collective behavior of social and living systems, such as pedestrian crowds, or cars on a freeway. Further use of biased ensembles in active matter models could therefore help establish ground-rules for their wider application in systems very far from equilibrium.

% ===============================================================================

\section*{DISCLOSURE STATEMENT}

The authors are not aware of any affiliations, memberships, funding, or financial holdings that might be perceived as affecting the objectivity of this review.

% -------------------------------------------------------------------------------

\section*{ACKNOWLEDGMENTS}

The authors acknowledge insightful discussions with
{\O}yvind L. Borthne,
Fernando Caballero,
Timothy Ekeh,
Yann-Edwin Keta,
Yuting Irene Li,
Tomer Markovich,
Cesare Nardini,
Takahiro Nemoto,
Patrick Pietzonka,
Sriram Ramaswamy,
Udo Seifert,
Thomas Speck,
Julien Tailleur,
Elsen Tjhung,
Laura Tociu,
Suriyanarayanan Vaikuntanathan,
and Fr\'ed\'eric van Wijland.
\'E.F. acknowledges support from an ATTRACT Grant of the FNR. M.E.C. is funded by the Royal Society. This work was funded in part by the European Research Council under the Horizon 2020 Programme, ERC grant agreement number 740269. This research was supported in part by the NSF (Grant No. NSF PHY-1748958)

% -------------------------------------------------------------------------------

\bibliographystyle{ar-style4.bst}
\bibliography{References}

\begin{thebibliography}{118}
\expandafter\ifx\csname natexlab\endcsname\relax\def\natexlab#1{#1}\fi

\bibitem{Marchetti2013}
Marchetti MC, Joanny JF, Ramaswamy S, et~al. 2013.
\textit{Rev. Mod. Phys.} 85:1143--1189

\bibitem{Bechinger2016}
Bechinger C, Di~Leonardo R, L\"owen H, Reichhardt C, et~al. 2016.
\textit{Rev. Mod. Phys.} 88:045006

\bibitem{Marchetti2018}
Fodor {\'E}, Marchetti MC. 2018.
\textit{Physica A} 504:106--120

\bibitem{Elgeti2015}
Elgeti J, Winkler RG, Gompper G. 2015.
\textit{Rep. Prog. Phys.} 78:056601

\bibitem{Ladoux2017}
Saw TB, Doostmohammadi A, Nier V, Kocgozlu L, Thampi S, et~al. 2017.
\textit{Nature} 544:212

\bibitem{Cavagna2014}
Cavagna A, Giardina I. 2014.
\textit{Annu. Rev. Condens. Matter Phys.} 5:183--207

\bibitem{Bartolo2019}
Bain N, Bartolo D. 2019.
\textit{Science} 363:46--49

\bibitem{Dauchot2010}
Deseigne J, Dauchot O, Chat\'e H. 2010.
\textit{Phys. Rev. Lett.} 105:098001

\bibitem{Palacci2013}
Palacci J, Sacanna S, Steinberg AP, Pine DJ, Chaikin PM. 2013.
\textit{Science} 339:936--940

\bibitem{Vicsek1995}
Vicsek T, Czir\'ok A, Ben-Jacob E, Cohen I, Shochet O. 1995.
\textit{Phys. Rev. Lett.} 75:1226--1229

\bibitem{Fily2012}
Fily Y, Marchetti MC. 2012.
\textit{Phys. Rev. Lett.} 108:235702

\bibitem{Toner1995}
Toner J, Tu Y. 1995.
\textit{Phys. Rev. Lett.} 75:4326--4329

\bibitem{Wittkowski2014}
Wittkowski R, Tiribocchi A, Stenhammar J, Allen RJ, et~al. 2014.
\textit{Nat. Commun.} 5:4351

\bibitem{Chate2020}
Chat\'e H. 2020.
\textit{Annu. Rev. Condens. Matter Phys.} 11:189--212

\bibitem{Cates2015}
Cates ME, Tailleur J. 2015.
\textit{Annu. Rev. Condens. Matter Phys.} 6:219--244

\bibitem{Tailleur2008}
Tailleur J, Cates ME. 2008.
\textit{Phys. Rev. Lett.} 100:218103

\bibitem{Maggi2015}
{Maggi} C, {Marini Bettolo Marconi} U, {Gnan} N, {Di Leonardo} R. 2015.
\textit{Sci. Rep.} 5:10742

\bibitem{Marchetti2014}
Yang X, Manning ML, Marchetti MC. 2014.
\textit{Soft Matter} 10:6477--6484

\bibitem{Brady2014}
Takatori SC, Yan W, Brady JF. 2014.
\textit{Phys. Rev. Lett.} 113:028103

\bibitem{Solon2015}
Solon AP, Fily Y, Baskaran A, Cates ME, Kafri Y, et~al. 2015.
\textit{Nat. Phys.} 11:673--678

\bibitem{Speck2015}
Bialk\'e J, Siebert JT, L\"owen H, Speck T. 2015.
\textit{Phys. Rev. Lett.} 115:098301

\bibitem{Zakine2020}
Zakine R, Zhao Y, Kne\ifmmode \check{z}\else
  \v{z}\fi{}evi\ifmmode~\acute{c}\else \'{c}\fi{} M, Daerr A, Kafri Y, et~al.
  2020.
\textit{Phys. Rev. Lett.} 124:248003

\bibitem{Paliwal2018}
Paliwal S, Rodenburg J, van Roij R, Dijkstra M. 2018.
\textit{New J. Phys.} 20:015003

\bibitem{Guioth2019}
Guioth J, Bertin E. 2019.
\textit{J. Chem. Phys.} 150:094108

\bibitem{Onsager1931}
Onsager L. 1931.
\textit{Phys. Rev.} 37:405--426

\bibitem{Kubo1966}
Kubo R. 1966.
\textit{Rep. Prog. Phys.} 29:255--284

\bibitem{Sekimoto1998}
Sekimoto K. 1998.
\textit{Prog. Theor. Phys. Supp.} 130:17--27

\bibitem{Seifert2012}
Seifert U. 2012.
\textit{Rep. Prog. Phys.} 75:126001

\bibitem{Maes1999gibbs}
Maes C. 1999.
\textit{J. Stat. Phys.} 95:367--392

\bibitem{Derrida2007}
Derrida B. 2007.
\textit{J. Stat. Mech.} 2007:P07023

\bibitem{Lecomte2007}
Lecomte V, Appert-Rolland C, van Wijland F. 2007.
\textit{J. Stat. Phys.} 127:51--106

\bibitem{Garrahan2007}
Garrahan JP, Jack RL, Lecomte V, Pitard E, et~al. 2007.
\textit{Phys. Rev. Lett.} 98:195702

\bibitem{Touchette2009}
Touchette H. 2009.
\textit{Phys. Rep.} 478:1--69

\bibitem{Jack2020}
Jack RL. 2020.
\textit{Eur. Phys. J. B} 93:74

\bibitem{Nardini2016}
Fodor {\'E}, Nardini C, Cates ME, Tailleur J, Visco P, et~al.
  2016{\natexlab{a}}.
\textit{Phys. Rev. Lett.} 117:038103

\bibitem{Dean1996}
Dean DS. 1996.
\textit{J. Phys. A: Math. Gen.} 29:L613

\bibitem{Tjhung2018}
Tjhung E, Nardini C, Cates ME. 2018.
\textit{Phys. Rev. X} 8:031080

\bibitem{Tiribocchi2015}
Tiribocchi A, Wittkowski R, Marenduzzo D, Cates ME. 2015.
\textit{Phys. Rev. Lett.} 115:188302

\bibitem{Nardini2017}
Nardini C, Fodor {\'E}, Tjhung E, van Wijland F, Tailleur J, et~al. 2017.
\textit{Phys. Rev. X} 7:021007

\bibitem{Lebowitz1999}
Lebowitz JL, Spohn H. 1999.
\textit{J. Stat. Phys.} 95:333--365

\bibitem{Mandal2017}
Mandal D, Klymko K, DeWeese MR. 2017.
\textit{Phys. Rev. Lett.} 119:258001

\bibitem{Puglisi2017}
Puglisi A, Marini Bettolo~Marconi U. 2017.
\textit{Entropy} 19:356

\bibitem{Speck2018}
Speck T. 2018.
\textit{{EPL}} 123:20007

\bibitem{Caprini2018}
Caprini L, Marconi UMB, Puglisi A, Vulpiani A. 2018.
\textit{Phys. Rev. Lett.} 121:139801

\bibitem{Shankar2018}
Shankar S, Marchetti MC. 2018.
\textit{Phys. Rev. E} 98:020604

\bibitem{Seifert2018}
Pietzonka P, Seifert U. 2018.
\textit{J. Phys. A: Math. Theor.} 51:01LT01

\bibitem{Ramaswamy2018}
Dadhichi LP, Maitra A, Ramaswamy S. 2018.
\textit{J. Stat. Mech.} 2018:123201

\bibitem{Bo2019}
Dabelow L, Bo S, Eichhorn R. 2019.
\textit{Phys. Rev. X} 9:021009

\bibitem{Borthne2020}
Borthne {\O}L, Fodor {\'{E}}, Cates ME. 2020.
\textit{New J. Phys.} 22:123012

\bibitem{Onsager1953}
Onsager L, Machlup S. 1953.
\textit{Phys. Rev.} 91:1505--1512

\bibitem{Caprini2019}
Caprini L, Marconi UMB, Puglisi A, Vulpiani A. 2019.
\textit{J. Stat. Mech.} 2019:053203

\bibitem{Martin2020b}
Martin D, de~Pirey TA. 2020 arXiv:2009.13476

\bibitem{Spinney2018}
Crosato E, Prokopenko M, Spinney RE. 2019.
\textit{Phys. Rev. E} 100:042613

\bibitem{Speck2016}
Speck T. 2016.
\textit{EPL} 114:30006

\bibitem{Solon2013}
Solon AP, Tailleur J. 2013.
\textit{Phys. Rev. Lett.} 111:078101

\bibitem{Alert2020}
Alert R, Joanny JF, Casademunt J. 2020.
\textit{Nat. Phys.} 16:682--688

\bibitem{Markovich2020}
Markovich T, Fodor {\'E}, Tjhung E, Cates ME. 2020 arXiv:2008.06735

\bibitem{Pietzonka2019}
Pietzonka P, Fodor {\'E}, Lohrmann C, Cates ME, Seifert U. 2019.
\textit{Phys. Rev. X} 9:041032

\bibitem{Ekeh2020}
Ekeh T, Cates ME, Fodor {\'E}. 2020.
\textit{Phys. Rev. E} 102:010101

\bibitem{Zakine2017}
Zakine R, Solon A, Gingrich T, van Wijland F. 2017.
\textit{Entropy} 19:193

\bibitem{Holubec2020}
Holubec V, Steffenoni S, Falasco G, Kroy K. 2020.
\textit{Phys. Rev. Research} 2:043262

\bibitem{Cates2021}
Fodor {\'E}, Cates ME. 2021 arXiv:2101.12646

\bibitem{Loos2020}
Loos SAM, Klapp SHL. 2020.
\textit{New J. Phys.} 22:123051

\bibitem{Kapral2018}
Gaspard P, Kapral R. 2018.
\textit{J. Chem. Phys.} 148:134104

\bibitem{Weber2019}
Weber CA, Zwicker D, Jülicher F, Lee CF. 2019.
\textit{Rep. Prog. in Phys.} 82:064601

\bibitem{Mazur}
Groot SRD, Mazur P. 1962.
{\it Non-Equilibrium Thermodynamics}.
Amsterdam: North-Holland

\bibitem{Kruse2004}
Kruse K, Joanny JF, J\"ulicher F, Prost J, Sekimoto K. 2004.
\textit{Phys. Rev. Lett.} 92:078101

\bibitem{Prost2017}
Prost J, J\"ulicher F, Joanny JF. 2015.
\textit{Nat. Phys.} 11:111--117

\bibitem{Martin2020}
Martin D, O'Byrne J, Cates ME, Fodor E, Nardini C, et~al. 2021.
\textit{Phys. Rev. E} 103:032607

\bibitem{Guo2015}
Fodor {\'{E}}, Guo M, Gov NS, Visco P, Weitz DA, van Wijland F. 2015.
\textit{{EPL}} 110:48005

\bibitem{Ahmed2018}
Ahmed WW, Fodor {\'E}, Almonacid M, Bussonnier M, et~al. 2018.
\textit{Biophys. J.} 114:1667--1679

\bibitem{Gnesotto2018}
Gnesotto FS, Mura F, Gladrow J, Broedersz CP. 2018.
\textit{Rep. Prog. Phys.} 81:066601

\bibitem{Sasa2005}
Harada T, Sasa Si. 2005.
\textit{Phys. Rev. Lett.} 95:130602

\bibitem{Szamel2019}
Szamel G. 2019.
\textit{Phys. Rev. E} 100:050603

\bibitem{Toyabe2010}
Toyabe S, Okamoto T, Watanabe-Nakayama T, et~al. 2010.
\textit{Phys. Rev. Lett.} 104:198103

\bibitem{Ahmed2016}
Fodor {\'{E}}, Ahmed WW, Almonacid M, Bussonnier M, Gov NS, et~al.
  2016{\natexlab{b}}.
\textit{{EPL}} 116:30008

\bibitem{Murrell2021}
Seara DS, Machta BB, Murrell MP. 2021.
\textit{Nat. Commun.} 12:392

\bibitem{Szamel2020b}
Flenner E, Szamel G. 2020.
\textit{Phys. Rev. E} 102:022607

\bibitem{Bo2020}
Dabelow L, Bo S, Eichhorn R. 2021.
\textit{J. Stat. Mech.} 2021:033216

\bibitem{Suri2020}
Fodor {\'E}, Nemoto T, Vaikuntanathan S. 2020.
\textit{New J. Phys.} 22:013052

\bibitem{Suri2019}
Tociu L, Fodor {\'E}, Nemoto T, Vaikuntanathan S. 2019.
\textit{Phys. Rev. X} 9:041026

\bibitem{Hansen}
Hansen JP, McDonald IR. 2013.
{\it Theory of Simple Liquids}.
Oxford: Academic Press

\bibitem{Suri2020b}
Tociu L, Rassolov G, Fodor {\'E}, Vaikuntanathan S. 2020 arXiv:2012.10441

\bibitem{Li2021}
Li YI, Cates ME. 2021.
\textit{J. Stat. Mech.} 2021:013211

\bibitem{Caballero2020}
Caballero F, Cates ME. 2020.
\textit{Phys. Rev. Lett.} 124:240604

\bibitem{Garrahan2009}
Garrahan JP, Jack RL, Lecomte V, Pitard E, et~al. 2009.
\textit{J. Phys. A} 42:075007

\bibitem{Bodineau2004}
Bodineau T, Derrida B. 2004.
\textit{Phys. Rev. Lett.} 92:180601

\bibitem{Appert2008}
Appert-Rolland C, Derrida B, Lecomte V, van Wijland F. 2008.
\textit{Phys. Rev. E} 78:021122

\bibitem{Jack2015}
Jack RL, Sollich P. 2015{\natexlab{a}}.
\textit{Eur. Phys. J. Special Topics} 224:2351--2367

\bibitem{Dolezal2019}
Dolezal J, Jack RL. 2019.
\textit{J. Stat. Mech.} 2019:123208

\bibitem{Hedges2009}
Hedges LO, Jack RL, Garrahan JP, Chandler D. 2009.
\textit{Science} 323:1309--1313

\bibitem{GarrahanLesanovsky2010}
Garrahan JP, Lesanovsky I. 2010.
\textit{Phys. Rev. Lett.} 104:160601

\bibitem{Weber2014}
Weber JK, Jack RL, Schwantes CR, Pande VS. 2014.
\textit{Biophys. J.} 107:974--982

\bibitem{Nemoto2019}
Nemoto T, Fodor {\'E}, Cates ME, Jack RL, Tailleur J. 2019.
\textit{Phys. Rev. E} 99:022605

\bibitem{Simha2008}
Simha A, Evans RML, Baule A. 2008.
\textit{Phys. Rev. E} 77:031117

\bibitem{Dill2013}
Press{\'{e}} S, Ghosh K, Lee J, Dill KA. 2013.
\textit{Rev. Mod. Phys.} 85:1115--1141

\bibitem{denH-book}
den Hollander F. 2000.
{\it Large deviations}.
Providence, RI: American Mathematical Society

\bibitem{bertsekas-book}
Bertsekas DP. 2005.
{\it Dynamic programming and optimal control}.
vol.~1.
Athena Scientific

\bibitem{dupuis-book}
Dupuis P, Ellis RS. 1997.
{\it A weak convergence approach to the theory of large deviations}.
New York: Wiley

\bibitem{Chetrite2015var}
Ch\'{e}trite R, Touchette H. 2015.
\textit{J. Stat. Mech.} 2015:P12001

\bibitem{Jack2015b}
Jack RL, Sollich P. 2015{\natexlab{b}}.
\textit{Eur. Phys. J.: Spec. Topics} 224:2351--2367

\bibitem{Chetrite2015}
Chetrite R, Touchette H. 2015.
\textit{Annales Henri Poincar{\'e}} 16:2005--2057

\bibitem{Jack2010}
Jack RL, Sollich P. 2010.
\textit{Prog. Theor. Phys. Supp.} 184:304--317

\bibitem{Touchette2016}
Tsobgni~Nyawo P, Touchette H. 2016.
\textit{Phys. Rev. E} 94:032101

\bibitem{Mallmin2020}
Cagnetta F, Mallmin E. 2020.
\textit{Phys. Rev. E} 101:022130

\bibitem{Keta2020}
Keta YE, Fodor E, van Wijland F, Cates ME, Jack RL. 2021.
\textit{Phys. Rev. E} 103:022603

\bibitem{Nemoto2016}
Nemoto T, Bouchet F, Jack RL, Lecomte V. 2016.
\textit{Phys. Rev. E} 93:062123

\bibitem{Nemoto2017first}
Nemoto T, Jack RL, Lecomte V. 2017.
\textit{Phys. Rev. Lett.} 118:115702

\bibitem{Limmer2018}
Ray U, Chan GKL, Limmer DT. 2018.
\textit{Phys. Rev. Lett.} 120:210602

\bibitem{Whitelam2019}
Jacobson D, Whitelam S. 2019.
\textit{Phys. Rev. E} 100:052139

\bibitem{Bertini2015}
Bertini L, De~Sole A, Gabrielli D, Jona-Lasinio G, et~al. 2015.
\textit{Rev. Mod. Phys.} 87:593--636

\bibitem{Whitelam2018}
Whitelam S, Klymko K, Mandal D. 2018.
\textit{J. Chem. Phys.} 148:154902

\bibitem{GrandPre2020}
GrandPre T, Klymko K, Mandadapu KK, Limmer DT. 2021.
\textit{Phys. Rev. E} 103:012613

\bibitem{Cagnetta2017}
Cagnetta F, Corberi F, Gonnella G, Suma A. 2017.
\textit{Phys. Rev. Lett.} 119:158002

\bibitem{GrandPre2018}
GrandPre T, Limmer DT. 2018.
\textit{Phys. Rev. E} 98:060601

\bibitem{Cagnetta2020}
Chiarantoni P, Cagnetta F, Corberi F, et~al. 2020.
\textit{J. Phys. A: Math. Theor.} 53:36LT02

\bibitem{Dolezal2020}
Dolezal J, Jack RL. 2020 arXiv:2012.05853

\bibitem{Horowitz2020}
Horowitz JM, Gingrich TR. 2020.
\textit{Nat. Phys.} 16:15--20

\end{thebibliography}

\end{document}